\documentclass[a4paper, reprint,twocolumn]{revtex4}
\usepackage{epsfig,graphicx}
\usepackage{amssymb,amsfonts,amsmath}
\allowdisplaybreaks

\begin{document}

\title{Towards a Theory of Additive Eigenvectors.}

\author{Sergei V. Krivov}
\affiliation{Astbury Center, Leeds University, Leeds, United Kingdom}
\email{s.krivov@leeds.ac.uk}

\begin{abstract}
The standard approach to solve stochastic equations is eigenvector decomposition. Using separation ansatz $P(i,t)=u(i)e^{\mu t}$ one obtains standard equation for eigenvectors $Ku=\mu u$, where $K$ is the rate matrix of the master equation. While universally accepted, the standard approach is not the only possibility. Using additive separation ansatz $S(i,t)=W(i)-\nu t$ one arrives at additive eigenvectors. Here we suggest a theory of such eigenvectors. We argue that additive eigenvectors describe conditioned Markov processes and derive corresponding equations. The formalism is applied to one-dimensional stochastic process corresponding to the telegraph equation. We derive differential equations for additive eigenvectors and explore their properties. The proposed theory of additive eigenvectors provides a new description of stochastic processes with peculiar properties.
\end{abstract}

\maketitle

\section{Introduction}
The standard approach to solve stochastic equations is eigenvector decomposition \cite{RiskenFokkerPlanckEquationMethods1984}. Consider, for example the master equation $\partial \vec{P}/\partial t=K\vec{P}$, where $\vec{P}=P(i,t)$ is the vector of probabilities, index $i$ denotes a state of the system, $t$ is time and $K$ is the rate matrix. Starting with the following separation ansatz $P(i,t)=u(i)e^{\mu t}$ one obtains standard equation $K\vec{u}_\mu=\mu \vec{u}_\mu$, where $\mu$ and $\vec{u}_{\mu}$ are an eigenvalue and the corresponding eigenvector \cite{RiskenFokkerPlanckEquationMethods1984}. The solution of the master equation is then represented as a linear combination  $P(i,t)=\sum_{\mu} u_{\mu}(i)e^{\mu t}$. The theory of eigenvectors and eigenvalues of such rate matrices is well developed. In particular, it is easy to show, that there is always eigenvalue $\mu=0$ with the corresponding eigenvector $\vec{u}_0$ representing equilibrium probabilities. All other eigenvalues have negative real part. For systems that obey the principle of detailed balance, the rate matrix can be made symmetric and all the other eigenvalues are real and negative. Thus, an arbitrary initial probability distribution exponentially relaxes to the limiting equilibrium distribution $\vec{u}_0$. The Fokker-Planck equation can be treated analogously \cite{RiskenFokkerPlanckEquationMethods1984}.

While universally accepted, the standard approach is not the only possibility. Recently we have started the development of a different formalism for solving stochastic equations \cite{KrivovMethoddescribestochastic2013}. The difference can be roughly summarized as the usage of different \textit{additive} separation ansatz $S(i,t)=W(i) -\nu t$, which results in solutions expressed by additive eigenvectors. The ansatz is not new and used, for example, in the analytical mechanics, where for solving the Hamilton-Jacobi equation, the Hamilton principal function is expressed as $S(q,\alpha, t)=W(q,\alpha) -E t$ \cite{LanczosVariationalPrinciplesMechanics1986}. However, the application of the additive ansatz to the description of stochastic dynamics has not been explored before. Here we suggest a theory of such eigenvectors. It provides a novel description of stochastic processes that leads to distinct new solutions with peculiar properties, very different from those of the standard approach.

\begin{figure}[htbp]
\centering
\includegraphics[width=0.8\linewidth]{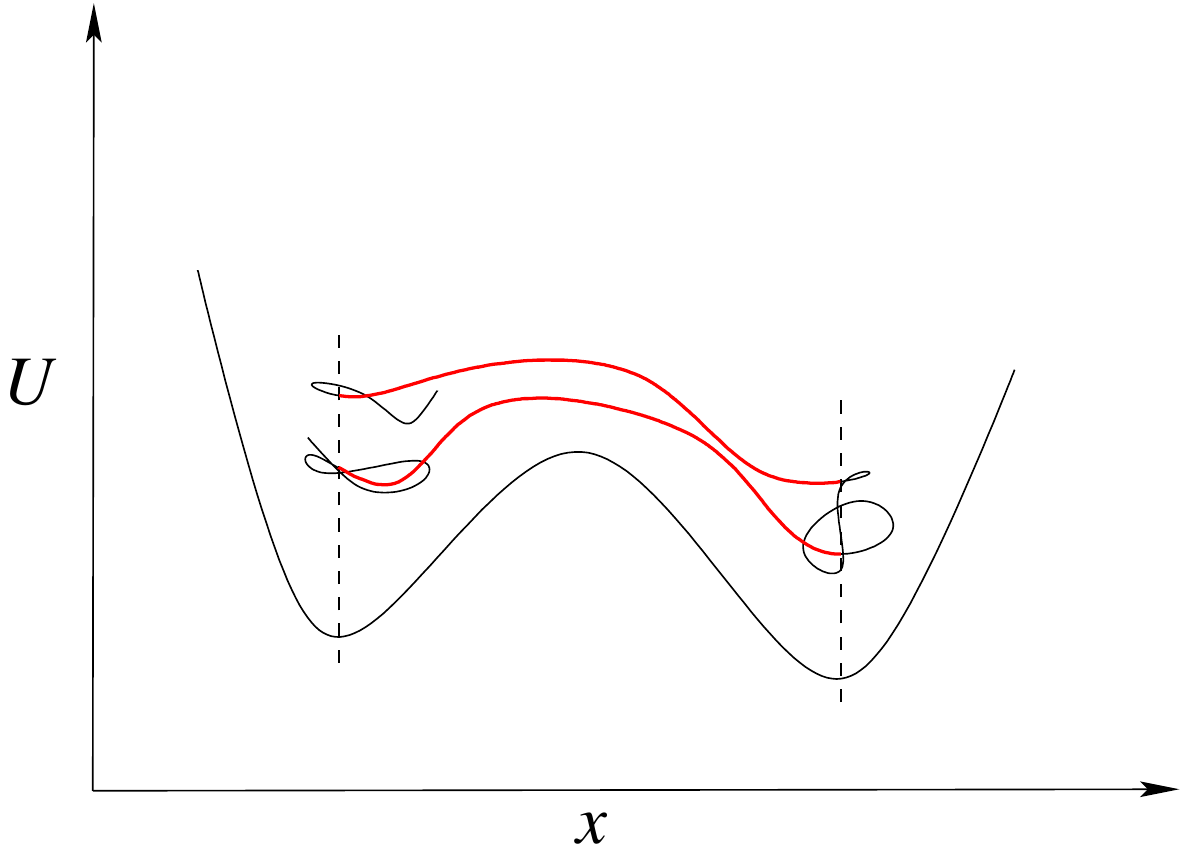}
\caption{A sketch of a fragment of a stationary trajectory on a model energy surface with two basins. Transition paths are shown by red color.}
\label{fig:twominima}
\end{figure}

One motivation for this work was the desire to decompose stochastic dynamics into a collection of stochastic eigenmodes - stationary stochastic processes periodic in some sense. Consider, for example, stationary stochastic diffusive dynamics on a model potential energy landscape with two basins shown on Fig. \ref{fig:twominima}. A free energy landscape of such form may describe a protein folding process \cite{BanushkinaOptimalreactioncoordinates2016}. Consider a long stationary trajectory of the system, where, in particular, the systems periodically visits the basins, e.g., a protein repeatedly folds and unfolds. On the trajectory one can identify so-called transition paths (TP) \cite{ETheoryTransitionPaths2006} - pieces of trajectory that start in one basin and end in the other basin, without returning to the first basin, shown by red color on Fig. \ref{fig:twominima}. The TPs are thought to describe the essence of the folding process, or, in general, the rare event of going over the barrier. They do not describe the dwelling in the basins. Aa consequence,  the mean folding time, for reasonably high free energy barriers, is much longer than the mean TP time. In other words, the folding process itself is quite fast, it is just very unlikely. The ensemble of TPs, considered as a long trajectory, describes stochastic dynamics of the system periodically visiting basins, with the mean period of two mean TP times. As the TP ensemble constitutes only a small and an atypical part of the equilibrium trajectory, its stationary probability distribution is different from that of the equilibrium trajectory. Such an ensemble of TPs can serve as an approximate example of what we mean by a stochastic eigenmode.

An element of arbitrariness in the definition of TPs is the criterion to decide when a trajectory has reached a basin. In this simple, one-dimensional case, we assumed that it is so when a trajectory has visited the bottom of a basin. However it is not clear how one can do it in general, e.g., in a single harmonic well, on a multidimensional potential energy surface with many minima or absolutely flat. One possible way to avoid this arbitrariness altogether, is to consider the ensemble of TPs not as a collection of paths between two states, but rather as an example of a stochastic periodic process or stochastic eigenmode. Considering all such eigenmodes one may hope that one of them would approximate the desired TP ensemble. Since TPs describe the essence of, e.g., the folding dynamics, stochastic eigenmodes could then provide a new type of optimal reaction coordinates \cite{BanushkinaOptimalreactioncoordinates2016}, which was another motivation for this work.

Assuming that decomposition into stochastic eigenmodes is possible, let us guess/discuss its expected properties.  Consider again long equilibrium trajectory on Fig. \ref{fig:twominima}. Assume that a stochastic eigenmode approximates the TPs. Consider the rest of the equilibrium trajectory without the TP ensemble. It describes stochastic dynamics inside basins, and if the formalism is general, should be described by another stochastic eigenmode. Generalizing, we may conclude that stationary stochastic dynamics can be decomposed into different ensembles of trajectories, where each ensemble describes stochastic periodic motion - a stochastic eigenmode. 

If the stationary dynamics is decomposed into different stochastic eigenmodes, it is reasonable to expect that stationary probability distribution of each eigenmode is computed based only on the trajectories in the eigenmode. In other words, each eigenmode has its own stationary distribution, which differs from the unique overall stationary probability distribution of the original stochastic dynamics.

This discussion suggests that dynamics in each eigenmode is described by a formalism different from the conventional Markov processes. The latter allows for only a unique stationary probability distribution.  Multiple stationary probability distributions are possible in the framework of conditioned Markov processes, which considers only a subset of allowed trajectories. For example, in TPs, out of all trajectories started in one basin, one keeps only those that reach the other basin first, i.e., the trajectories  are conditioned to start in one basin and to end in the other. Analogously, trajectories in a stochastic eigenmode can be conditioned to start and to end in the eigenmode. In other words, each eigenmode contains only a sub-ensemble of the entire equilibrium ensemble of trajectories. Such a conditioning of Markov dynamics on both the past and the future is an essential part of the description and is known as conditioned or reciprocal Markov processes \cite{Schrodinger1931, Wakolbinger1992, ColletQuasiStationaryDistributionsMarkov2013}. Also, stationary trajectories can be straightforwardly reversed in time. It means that such a description should treat forward and time-reversed stationary dynamics on equal footing, which is a feature of conditioned Markov processes.

In summary, we seek a formalism, which decomposes a stationary stochastic dynamics into a collection of periodic stochastic stationary processes, or stochastic eigenmodes, each with its own stationary probability distribution. We decompose a conventional, conditioning-free Markov process into a collection of conditioned Markov processes, conditioned on being periodic. An aim of this work is to show that such a decomposition can be obtained with additive eigenvectors.

The paper is as follows. We start by introducing additive eigenvectors in a familiar setting of conventional Markov processes. We then extend the formalism by considering additive eigenvectors as conditioned Markov processes. We use the theory of quasi-stationary distributions (QSD) \cite{ColletQuasiStationaryDistributionsMarkov2013} to derive a set of equations that describe additive eigenvectors as conditioned Markov processes for a Markov chain. We illustrate the formalism by applying it to the stochastic process corresponding to the telegraph equation \cite{Kacstochasticmodelrelated1974}, which, here forth is referred as telegraph process for brevity. Intriguingly, the obtained equation, while describing a classical stochastic process, appears to be very similar to the one-dimensional relativistic quantum Dirac equation. We present illustrative solutions of the derived equation and compare them with solutions of  the Dirac equation and with numerical simulations of conditioned telegraph trajectories. We end with a concluding discussion. 

\section{Additive eigenvectors for a Markov chain}
\label{sect:addevstd}
In this section we introduce the formalism of additive eigenvectors as conditioned Markov processes and derive the corresponding equation for a Markov chain. 
We start with a simple model system and introduce additive eigenvectors in a familiar case of conventional Markov processes. We then consider additive eigenvectors as conditioned Markov processes and derive corresponding equations by using the QSD theory \cite{ColletQuasiStationaryDistributionsMarkov2013}.

\subsection{Additive eigenvectors for forward dynamics}
Consider a system with $N$ states on a circle, where transitions are allowed only to the right ($j\rightarrow j+1$) with rate $r$. Stochastic dynamics is described by the following rate or master equation (transition $N-1 \rightarrow N$ should read $N-1 \rightarrow 0$)
\begin{align}
\partial P(j,t)/\partial t&= -r P(j,t)+r P(j-1,t).
\label{exmpl2}
\end{align}
or, in the limit of small $\Delta t$, by the following Markov chain 
\begin{align}
P(j,t+\Delta t)=(1-r \Delta t)P(j,t)+r \Delta tP(j-1,t).
\label{exmpl}
\end{align}

Using the multiplicative ansatz $P(j,t)=u(j)e^{\mu t}$, one finds $N$ standard eigenvalues $\mu_k=re^{-ik}-r$ and (right) eigenvectors $u_k(j)=e^{ikj}$, numbered by $k$, where $k=2\pi l/N$ for $l=0,1,...,N-1$. Combining with complex-conjugate eigenvectors one finds real solutions in the form 
\begin{subequations}\label{stdeig}
\begin{align}
&P^c_k(j,t)=e^{r(\cos(k)-1)t}\cos[kj-r\sin(k)t]\\
&P^s_k(j,t)=e^{r(\cos(k)-1)t}\sin[kj-r\sin(k)t],
\end{align}
\end{subequations}
which describe waves moving to the right and exponentially decaying with time. The general solution can be written as 
\begin{align}
P(j,t)=\sum_k a_k P^c_k(j,t)+b_kP^s_k(j,t),
\end{align}
where  $a_k$ and $b_k$ are real. The stationary probability distribution is then $P^{st}(j)=P^c_0(j,t)=1$.

Using the additive ansatz, we seek a solution in the form $\theta(j,t)=u(j)(W(j)-\nu t)$, where $u(j)$ and $W(j)$ are some, so far unknown, vectors. Substituting $\theta(j,t)$ instead of $P(j,t)$ into Eqs. \ref{exmpl} one obtains
\begin{align} \label{addeig}
u(j)[W(j)-&\nu(t+\Delta t)]=(1-r \Delta t) u(j)[W(j)-\nu t]\nonumber\\
                         &+r \Delta tu(j-1)[W(j-1)-\nu t]
\end{align}
By requiring Eq. \ref{addeig} to be valid for all $t$ one obtains two equations
\begin{subequations}\label{addevexmpl}
\begin{align}
&u(j)=(1-r \Delta t)u(j)+r \Delta tu(j-1) \\
&u(j)(W(j)-\nu \Delta t)= \nonumber\\&(1-r \Delta t)u(j)W(j)+r \Delta tu(j-1)W(j-1) 
\end{align}
\end{subequations}
The first equation, describing the cancellation of terms proportional to $t$, is the equation on the stationary probabilities, thus $u(j)=P^{st}(j)=1$. The second equation describes additive eigenvector $W(j)$ with additive eigenvalue $\nu$. The reason for this terminology will become clear later. From the second equation one finds $W(j)-W(j-1)=\nu/r$. Since this equation defines $W(j)$ and $\nu$ up to an overall factor, we can set $\nu=r$ and obtain the following solution $\theta(j,t)=j-rt$. 

Additive eigenvectors are multivalued functions. Equation $W(j)-W(j-1)=\nu/r$ suggests that $W(j)$ grows linearly with $j$ and has a general solution $W(j)=W(0)+j\nu/r$. However, for $j=N$ one obtains that $W(0)=W(N-1)+\nu/r$. It either means that $\nu=0$ and $W(j)=W(0)$ and one obtains the standard single-valued solution, or that $W(j)$ is a multivalued function, that changes by $N\nu/r$ when the system revisits any state by completing the trip around the entire circle. It is analogous to the multivalued function $\ln z$ where $z$ is a complex number. If one performs an analytic continuation moving around $z=0$ in the complex plane, e.g., along the unit circle $|z|=1$, and returns to the same point, $\ln z$ increments by $2\pi i$. It is also analogous  to the angle variable $\phi$, which increments by $2\pi$ after each revolution. We show in Appendix \ref{propev} that additive eigenvectors are multivalued functions for Markov chains that obey the detailed balance condition or that have a finite number of states.

We have obtained a new solution of Eqs. \ref{exmpl} or \ref{exmpl2}. This solution can not be expressed as a linear combination of standard solutions $P^c_k$ and $P^s_k$ because they decay exponentially with time. There is no contradiction, as the new solution belongs to a different functional class of multivalued functions. What is the interpretation of the new solution $\theta(j,t)$? It can not be interpreted directly as a probability distribution. It can be interpreted analogously to a stationary wave function. The solution has a generic form of $\theta(j,t)=u(j)(W(j)-\nu t)$. The single valued part $u(j)$ can be interpreted as a stationary probability distribution, while the multivalued part $W (j)$ can be interpreted as a phase, which describes time evolution of (stationary) stochastic dynamics.

Seeking solution in the form $\theta(j,t)=u(j)e^{\mu t}(W(j)-\nu t)$, it is straightforward to show that for every standard eigenvector $u(j)$ and eigenvalue $\mu$ one can find corresponding additive eigenvector $W(j)$ and eigenvalue $\nu$. However we limit our discussion to stationary solutions only, since our am is to decompose the dynamics into a collection of stationary stochastic processes. 

For a generic Markov chain
\begin{equation}\label{mc}
P(i,t+\Delta t)=\sum_j P_{\Delta t}(i|j)P(j,t),
\end{equation}
where $P(j,t)$ is the probability in state $j$ at time $t$ and $P_{\Delta t}(i|j)=P(i,t+\Delta t|j,t)$ is the stationary transition probability from state $j$ to $i$ after time interval $\Delta t$, seeking solution in the form $\theta(j,t)=u(j)(W(j)-\nu t)$ 
one obtains 
\begin{subequations}\label{addevgen}
\begin{align}
&u(i)=\sum_j P_{\Delta t}(i|j)u(j)\\
&u(i)(W(i)-\nu \Delta t)=\sum_j P_{\Delta t}(i|j)u(j)W(j) \label{addevgen:b}
\end{align}
\end{subequations}
The first equation shows that $u(j)$ is a standard eigenvector representing stationary probability distribution $P^{st}(j)$. The second equation defines corresponding $W(j)$ and $\nu$. 

\subsection{Time-reversal of stationary dynamics}
Time reversal of stationary dynamics is again stationary dynamics. As mentioned in introduction, we require description that is symmetric with respect to the past and the future, meaning that an additive eigenvector should describe both the forward and time-reversed dynamics.

Let $X(k\Delta t)$ for $0 \le k \le L$ denote the forward trajectory, then the time-reversed trajectory is defined as $\hat{X}(k \Delta t)=X((L-k) \Delta t)$, here and below hats denote quantities of time-reversed dynamics. Let $n(i,t+\Delta t|j,t)=P(i,t+\Delta t|j,t)P^{st}(j,t)$ denote the number of transitions from state $j$ to state $i$ after time interval $\Delta t$. While $n(i,t+\Delta t|j,t)$, $P(i,t+\Delta t|j,t)$ and $P^{st}(j,t)$ are stationary quantities and do not depend on $t$, we employ formal time dependence as a convenient bookkeeping device. For the time-reversed trajectory one has  $\hat{n}(j, t|i,t+\Delta t)=n(i,t+\Delta t| j, t)$ and $\hat{P}^{st}(i,t)=P^{st}(i,t)$, and hence $\hat{P}(j,t|i,t+\Delta t)=P(i,t+\Delta t|j,t)P^{st}(j,t)/P^{st}(i,t+\Delta t)$. The formal time dependence of $\hat{P}(j,t|i,t+\Delta t)$ indicates that transitions happen from $t+\Delta t$ to $t$. Equation for additive eigenvector for time-reversed dynamics is
\begin{align}
&P^{st}(j,t)W(j)= \nonumber\\&\sum_i \hat{P}(j,t|i,t+\Delta t)P^{st}(i,t+\Delta t)(W(i)-\nu \Delta t), \label{addevgenbw}
\end{align}
which can also be written, by using transition probabilities for forward dynamics, as
\begin{align}
\sum_i P(i,t+\Delta t|j,t)W(i)=W(j)+\nu \Delta t. \label{addevgenbw2}
\end{align}
The equation explains why we called $W(j)$ and $\nu$ additive eigenvector and eigenvalue, correspondingly. The action of the transition probability matrix does not change the eigenvector and only shifts it by an additive constant.

For the model system described by Eq. \ref{exmpl} one obtains
\begin{align}
r \Delta t W(j+1)+(1-r \Delta t)W(j)=W(j)+\nu \Delta t, \label{addevbw}
\end{align}
which leads to $W(j+1)-W(j)=\nu/r$. Thus $\theta(j,t)=j-rt$ describes both forward and time-reversed dynamics.

Note however, that for a bit more complex system with state dependent rate $r_j$, one obtains $W(j)-W(j-1)=\nu/r_{j}$ and $W(j+1)-W(j)=\nu/r_{j}$ for forward and time-reversed dynamics, respectively. This means that it is not possible to find an additive eigenvector that describes both forward and time-reversed dynamics, unless  
all $r_j$ are equal. The extension of the formalism necessary for such systems is discussed in the following section.

Both Eqs. \ref{addevgen:b} and \ref{addevgenbw2} can be put in the following form
\begin{subequations}\label{addevgenall}
\begin{align}
&\sum_j \hat{P}(j,t|i,t+\Delta t)[S(i,t+\Delta t)-S(j,t)]=0\\
&\sum_i P(i,t+\Delta t|j,t)[S(i,t+\Delta t)-S(j,t)]=0,
\end{align}
\end{subequations}
where we introduced $S(i,t)=W(i) -\nu t$. The equations describe that average change of the phase $S$ along the forward and time-reversed trajectories is zero.


Using $P_{\Delta t}(i|j)=e^{K(i|j)\Delta t}\approx \delta_{ij}+ K(i|j)\Delta t$, where $K(i|j)$ is the reaction rate from $j$ to $i$ and $\delta_{ij}$ is the Kronecker $\delta$-function, one obtains for $\Delta t \rightarrow 0$
\begin{subequations}\label{addevK}
\begin{align}
&\sum_j \hat{K}(j|i)(S(i,t)-S(j,t))=-\partial S(i,t)/\partial t\\
&\sum_i K(i|j)(S(i,t)-S(j,t))=-\partial S(j,t)/\partial t,
\end{align}
\end{subequations}
here $\hat{K}(j|i)=K(i|j)P^{st}(j)/P^{st}(i)$ are the reaction rates for the time-reversed dynamics.

\subsection{Additive eigenvectors as conditioned Markov processes.}
\label{sect:addevcond}
The formalism presented in the previous sections provides only a single stationary
additive eigenvector solution and correspondingly a unique stationary probability distribution. However, as was mentioned in the introduction we are looking for a formalism that can be used to decompose stationary dynamics into a number of different eigenmodes. Each eigenmode contains a sub-ensemble of the entire equilibrium ensemble of trajectories, which describes a stationary stochastic process with a different stationary probability distribution. It means that when a set of trajectories leave a particular state, we consider only those trajectories, which stay in the sub-ensemble. Such Markov processes, conditioned on the future can be described by the formalism of QSD \cite{ColletQuasiStationaryDistributionsMarkov2013}.

Consider a Markov chain with rate matrix $K^0(i|j)$ that describes process $X$. We want to specify/describe a sub-ensemble of trajectories that form a stationary stochastic process $Y$ which is periodic in some sense. Since we consider just a sub-ensemble of trajectories, and not the entire equilibrium ensemble of trajectories, it means that some of the trajectories that leave current state $i$ leave also the sub-ensemble. Let $\chi(i)$ denote the rate of leaving current sub-ensemble from state $i$. Process $Y$ consists of stochastic trajectories that never leave the sub-ensemble. Its description is provided by the formalism of QSD \cite{ColletQuasiStationaryDistributionsMarkov2013}. Let $K^A(i|j)=K^0(i|j)-\delta_{ij}\chi(i)$ and let $\theta>0$, $q(i)>0$ and $v(i)>0$ denote the Perron-Frobenius eigenvalue and the corresponding left and right eigenvectors of $K^A(i|j)$

\begin{subequations}\label{pf}
\begin{align}
&\sum_i q(i) K^A(i|j)=-\theta q(j)\\
&\sum_j K^A(i|j)v(j)=-\theta v(i)
\end{align}
\end{subequations}
Then, process $Y$ is again a Markov process with reaction rate matrix 
\begin{align}\label{kij}
K(i|j)=K^A(i|j)q(i)/q(j)+\delta_{ij} \theta
\end{align}
and quasi-stationary probability distribution $P^{st}(j)=q(j)v(j)$. We drop prefix \textit{quasi} henceforth. Namely, one can straightforwardly verify that 
\begin{subequations}\label{qsd}
\begin{align}
&\sum_i K(i|j)=0\\
&\sum_j K(i|j)P^{st}(j)=0
\end{align}
\end{subequations}
Since process Y is a standard Markov process, one can use Eqs. \ref{addevK} to find corresponding additive eigenvectors for forward and time-reversed dynamics. The rate matrix for the time-reversed dynamics is $\hat{K}(j|i)=K(i|j)P^{st}(j)/P^{st}(i)=K^A(i|j)v(j)/v(i)+\delta_{ij} \theta$. Thus, for an arbitrary vector $\chi(i)$ one can find a corresponding stationary process $Y$ with the corresponding stationary distribution $P^{st}(j)$ and the corresponding additive eigenvectors for forward and time-reversed dynamics.
The additive eigenvectors for forward and time-reversed dynamics are not necessarily equal. Next we impose a constraint and consider only those $\chi(i)$ for which additive eigenvectors for forward and time-reversed dynamics are the same. We call such a process $Y$ conditioned on being an additive eigenvector.

Note that the results do not change if one shifts both $\chi(i)$ and $\theta$ by the same constant, i.e., $\chi(i)\rightarrow \chi(i)-a$ and $\theta \rightarrow \theta -a$. In particular, one may set $\theta=0$. In that case some of $\chi(i)$ become negative. They have the following interpretation. The positive $\chi(i)$ describe loss of trajectories from the sub-ensemble of trajectories corresponding to current additive eigenvector or more briefly, loss of trajectories from the current additive eigenvector, to other additive eigenvectors. Negative $\chi(i)$ then describe influx of trajectories from other additive eigenvectors to the current additive eigenvector. This influx is such that it does not alter $q$ and $v$ eigenvectors, and thus the corresponding properties of the conditioned process $Y$, in particular the rate matrix $K(i,j)$ and the stationary probabilities $P^{st}(j)$. For $\theta=0$ the loss of trajectories to and influx of trajectories from other additive eigenvectors balance out. In this case the equations can be summarized as 

\begin{subequations}\label{condaddev}
\begin{align}
&\sum_i q(i) (K^0(i|j)-\chi(i)\delta_{ij})=0\\
&\sum_j (K^0(i|j)-\chi(i)\delta_{ij})v(j)=0\\
&\sum_i K^0(i|j)q(i)/q(j)[W(i)-W(j)]=\nu\\
&\sum_j K^0(i|j)v(j)/v(i)[W(i)-W(j)]=\nu
\end{align}
\end{subequations}
By eliminating  $\chi(i)$ from the first two equations, 
one obtains equation in the form
\begin{align}
\label{condP}
\sum_{j, j\ne i} &K^0(i|j)q(i)/q(j) P^{st}(j)- \sum_{j, j\ne i} K^0(j|i)q(j)/q(i)P^{st}(i) \nonumber \\
&=0
\end{align}
Eqs. \ref{condP} and \ref{condaddev}c,d constitute a closed system of equations, that describes additive eigenvectors without any reference to $\chi(i)$, the exact mechanism of exchange between the sub-ensembles of trajectories, corresponding to different additive eigenvectors.

The derived equations (Eqs. \ref{condaddev} or Eqs. \ref{condP} and \ref{condaddev}c,d) describe a conditioned Markov process. That process is conditioned to be stationary and periodic in the sense that it is described by an additive eigenvector which describes an advancement of the phase. It is conditioned to have the same additive eigenvectors for forward and time-reversed dynamics. We call such a process Markov process conditioned on being an  additive eigenvector. This conditioned process describes a sub-ensemble of trajectories of the entire stationary ensemble of trajectories of the process $X$. $\chi(i)$ is the rate with which the sub-ensemble of trajectories corresponding to the current additive eigenvector exchanges trajectories with other additive eigenvectors. $q(i)$ is the probability that a trajectory, starting from state $i$ will stay in the sub-ensemble, $v(i)$ is the probability to be at state $i$ for the sub-ensemble of trajectories. Correspondingly, $q(i)v(i)$ is the probability that trajectory starting and staying in the sub-ensemble will by found at state $i$, i.e., the stationary distribution for this sub-ensemble of trajectories. One can make the description more symmetric to the time reversal by introducing probability that a trajectory found at $i$ came from the sub-ensemble $\hat{q}(i)=v(i)/P^0(i)$, where $P^0(i)$ is the stationary probability for process $X$.  In this case the stationary probability for the sub-ensemble $P^{st}(i)=P^0(i)q(i)\hat{q}(i)$ equals the conditioning-free probability, times the probability that outgoing trajectory will stay in the sub-ensemble, times the probability that incoming trajectory came from the sub-ensemble. The sub-ensemble of trajectories is conditioned on both the past and the future.

In the derivation of the equations (Eqs. \ref{condaddev} and \ref{condP}) we assumed that quantities $\chi(i)$, $q(i)$, $v(i)$ (and correspondingly $P^{st}(i)$) and $W(i)$  are time independent as it should be for an additive eigenvector. However, one may postulate that the description is generally valid for time dependent solutions, e.g., describing a superposition of additive eigenvectors. In this case all these quantities become time dependent. In particular, Eq. \ref{condP}, which is the balance equation for $P^{st}(i)$, equals $\partial P^{st}(i)/\partial t$. The equations take the form (cf. Eqs. \ref{addevK}, we dropped the superscript st from $P^{st}$ and the time variables)
\begin{subequations}\label{condaddevt}
\begin{align}
&P(i)=q(i)v(i)\\
&\sum_i K^0(i|j)q(i)/q(j)[S(i)-S(j)]=-\partial S(j)/\partial t\\
&\sum_j K^0(i|j)v(j)/v(i)[S(i)-S(j)]=-\partial S(i)/\partial t\\
&\sum_{j, j\ne i} K^0(i|j)q(i)/q(j) P(j)- \sum_{j, j\ne i} K^0(j|i)q(j)/q(i)P(i) \nonumber \\
&=\partial P(i)/\partial t
\end{align}
\end{subequations}

\section{Application to the telegraph process}
\subsection{Additive eigenvectors for the telegraph process}
In this section the developed formalism is illustrated on the stochastic telegraph process \cite{Kacstochasticmodelrelated1974}. It describes a particle that moves with constant speed ($c$) in one dimension, and reverses direction with rate $r$. The time evolution of probability distributions $P_1(x,t)$ and $P_2(x,t)$ for particle moving to the right and left, respectively, is given by 
\begin{subequations} \label{tlg:all}
\begin{align}
\partial P_1/ \partial t_+&= -r P_1 +r P_2\\
\partial P_2/ \partial t_-&= -r P_2 + r P_1
\end{align}
\end{subequations}
where $\partial /\partial t_+=\partial /\partial t+c\partial /\partial x$ and  $\partial /\partial t_-=\partial /\partial t-c\partial /\partial x$.

The multivalued character of additive eigenvectors can be specified rather 
straightforwardly. The only apparent possibility, which respects the translational symmetry, is to change the branch of a multivalued function, when the particle changes the direction of movement. Namely, one uses two functions (two branches) $W_1(x)$ and $W_2(x)$ which describe the motion to the right and to the left, respectively. 


The derived equations (e.g., Eqs. \ref{condaddevt}) can not be applied directly to the telegraph process as it is not a Markov chain in the strict sense, since it contains pieces of deterministic motion. However it can be straightforwardly approximated by a Markov chain. For example, the deterministic movement to the right, with speed $c$,
can be approximate in the limit of $\Delta x \rightarrow 0$ with Markov chain, where transition are allowed only to the right neighbor with rate $K((i+1)\Delta x|i \Delta x)=c/\Delta x$.    

Approximating the telegraph process by a Markov chain, one obtains the following equations (a derivation is provided in the Appendix \ref{dertlg})
\begin{subequations} \label{tlgladdcond:all}
\begin{align}
&P_i=q_iv_i\\
&\partial P_1/ \partial t_+=-rq_2/q_1 P_1 +rq_1/q_2 P_2\\
&\partial P_2/ \partial t_-=-rq_1/q_2 P_2 + rq_2/q_1 P_1\\
&\partial S_1/\partial t_+ +rq_2/q_1[S_2-S_1]=0\\
&\partial S_2/\partial t_- +rq_1/q_2[S_1-S_2]=0\\
&\partial S_1/\partial t_+ +rv_2/v_1[S_1-S_2]=0\\
&\partial S_2/\partial t_- +rv_1/v_2[S_2-S_1]=0,
\end{align}
\end{subequations}
here $S_i(x,t)=W_i(x)-\nu t$. Variables $P_i$, $q_i$ and $v_i$ are functions of $x$ and $t$.  Subscripts 1 and 2 denote quantities describing movement to the right and left, correspondingly. For a solution describing a single additive eigenvector, $P_i$, $q_i$, $v_i$ and $W_i$ are time independent. For a general case, e.g., a solution describing a superposition of additive eigenvectors, these quantities are time dependent.

\begin{figure}[htbp]
\centering
\includegraphics[width=0.6\linewidth]{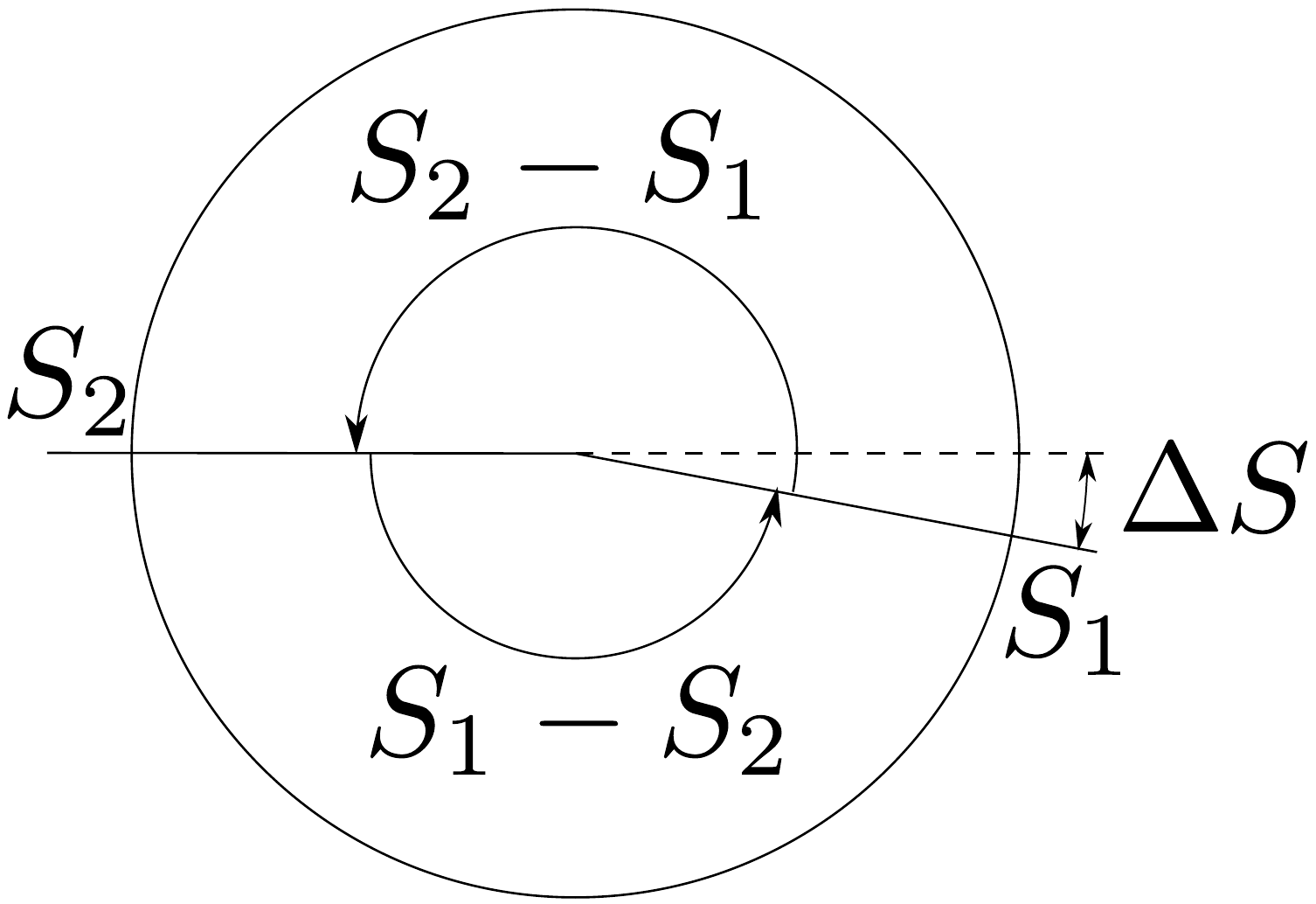}
\caption{$S_1$ and $S_2$ are analogous to angles. Differences $S_2-S_1$ and $S_1-S_2$ equal to the corresponding angle differences in radians, measured counter clockwise and divided by $\pi$. }
\label{fig:ds}
\end{figure}

Functions  $S_1(x,t)$ and $S_2(x,t)$ are multivalued. It means that going from $S_1$ to $S_2$ and then from $S_2$ back to $S_1$ should increment the value of $S_1$. An intuitive way to understand that is to consider $S_1$ and $S_2$ as rescaled angles (Fig. \ref{fig:ds}). Place points $S_1$ and $S_2$ on the circle and measure angle difference in radians always going counter clockwise. Let us denote the differences $S_2 -S_1=\Delta S_1$ and $S_1 -S_2=\Delta S_2$, then multivaluedness means that $\Delta S_1+\Delta S_2\ne 0$. Impose normalization that 
$\Delta S_1+\Delta S_2=2$. Let $2\Delta S=\Delta S_1- \Delta S_2$, then
\begin{subequations} \label{aa5:all}
\begin{align}
&S_2 -S_1=1+\Delta S\\
&S_1-S_2=1-\Delta S
\end{align}
\end{subequations}
When $\Delta S=0$, $\Delta S_1= \Delta S_2$ and the points are opposite to each other, i.e., $\Delta S$ measures the distance between a point and the opposite image of the other point. 

Using Eqs. \ref{tlgladdcond:all}a,d,f one finds that
\begin{subequations} \label{aa4:all}
\begin{align}
q_2/q_1&=R_2/R_1\sqrt{(1-\Delta S)/(1+\Delta S)}\\
v_2/v_1&=R_2/R_1\sqrt{(1+\Delta S)/(1-\Delta S)}
\end{align}
\end{subequations}
Substituting into Eqs. \ref{tlgladdcond:all} one obtains four equations with four unknowns $R_i(x,t)$ and $S_i(x,t)$
\begin{subequations} \label{fin:all}
\begin{align}
&\partial S_1/\partial t_+ +rR_2/R_1\sqrt{1-\Delta S^2}=0\\
&\partial S_2/\partial t_- +rR_1/R_2\sqrt{1-\Delta S^2}=0\\
&\frac{1}{rR_2}\frac{\partial R_1}{\partial t_+}=-\frac{1}{rR_1}\frac{\partial R_2}{\partial t_-}=\Delta S/\sqrt{1-\Delta S^2}\label{fin:c}
\end{align}
\end{subequations}

For the telegraph process with varying rates $r_1(x)$ and $r_2(x)$, to turn left and right,  respectively, one again obtains Eqs. \ref{fin:all}, with $r=\sqrt{r_1 r_2}$.

A  brief, preliminary analysis of the solutions of the derived equations will be performed in the next section. Here we just note that Eqs. \ref{fin:all}, while describing a classic stochastic process, are very similar to the one-dimensional relativistic quantum-mechanical Dirac equation, and for some values of the parameters they have very similar solutions. Consider the one-dimensional Dirac equation in the following representation  
\begin{subequations} \label{dirac0:all}
\begin{align}
\partial \psi_1/\partial t_+&=ir\psi_2\\
\partial \psi_2/\partial t_-&=ir\psi_1,
\end{align}
\end{subequations}
here $r=mc^2/\hbar$. If one rewrites the equation in the form that separates the amplitudes and phases of the wavefunction $\psi_j=R_je^{iS_j}$ one obtains 
\begin{subequations} \label{dirac2:all} 
\begin{align}
&\partial S_1/\partial t_+ +rR_2/R_1\cos(\Delta S)=0 \label{dirac1:a}\\
&\partial S_2/\partial t_- +rR_1/R_2\cos(\Delta S)=0\\
&\frac{1}{rR_2}\frac{\partial R_1}{\partial t_+}=-\frac{1}{rR_1}\frac{\partial R_2}{\partial t_-}=\sin(\Delta S), \label{dirac1:c}
\end{align}
\end{subequations}
where $S_2-S_1=\pi+\Delta S$. In the regime of small $\Delta S$, when $\cos(\Delta S)\sim \sqrt{1-\Delta S^2}+ O(\Delta S^4)$ and $\sin(\Delta S)\sim \Delta S/\sqrt{1-\Delta S^2}+ O(\Delta S^3)$, one expects similar solutions. In particular, both equations have the following position independent solutions:
$S_1=kx-\nu t$, $\Delta S=0$, $R_1=\sqrt{1+ck/\nu}$, and $R_2=\sqrt{1-ck/\nu}$, where $\nu^2=r^2+c^2k^2$.

How the additive eigenvector conditioning of a classical stochastic process can lead to equations that obey Lorentz transformations? The conditioning leads to the rescaling of rates of the telegraph equation (Eqs. \ref{tlgladdcond:all}b,c). This rescaling is equivalent to the rescaling obtained due to Lorentz transformations. We demonstrate this in the Appendix \ref{derlor}, where we also show how Eqs. \ref{fin:all} can be derived by acting with global and local Lorentz transformations on the equations describing the solution in the rest system (i.e., $k=0$).

\subsection{Illustrative solutions of Eqs. \ref{fin:all} and \ref{dirac2:all}}
A thorough analysis of the properties of the derived equations Eqs. \ref{fin:all} will be done elsewhere. Here we just illustrate a few solutions of the derived equations  and compare them with the standard eigenfunctions of the Dirac equations Eqs. \ref{dirac2:all}.

\begin{figure}[h]
\centering
\includegraphics[width=0.7\linewidth]{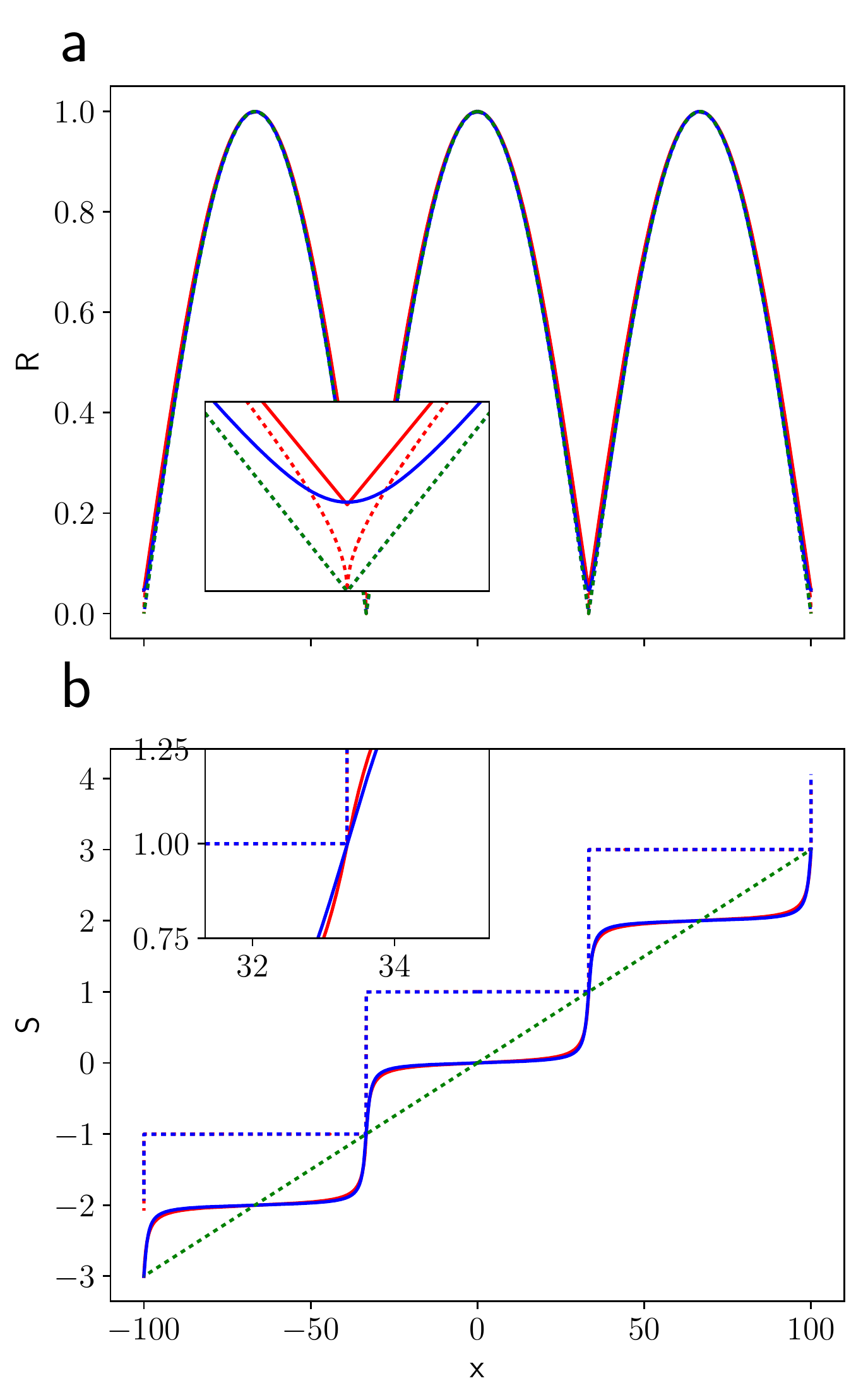}
\caption{3rd eigenfunctions of Eqs. \ref{fin:all} and \ref{dirac2:all} for $L=100$, corresponding to the non-relativistic regime. Panel a) shows $R_1$  and $R_2$ of Eqs. \ref{fin:all} by solid and dotted red lines, respectively. $R_1$  and $R_2$ of Eqs. \ref{dirac2:all} are shown by solid and dotted blue lines, respectively. Non relativistic  eigenfunction $\cos(kx)$ is shown by green dotted line. Panel b) shows $S_1$ and $S_2$ of Eqs. \ref{fin:all} by solid and dotted red lines, respectively and 
$S_1/(\pi/2)$ and $S_2/(\pi/2)-1$ of Eqs. \ref{dirac2:all} by solid and dotted blue lines, respectively. The green dotted line shows $kx$. The insets show the magnified view of the region around the third local minimum.}
\label{fig:bound100}
\end{figure}

Consider a setup that mimics a standing wave in a box with infinite walls at $x=\pm L$; we set $c=1$ and $r=1$. We first present solutions where $|\Delta S|<1$ for Eqs. \ref{fin:all} and  $|\Delta S|<\pi$ for the Dirac equation. The eigenfunctions were determined by the shooting method, by starting from $x=0$ and continuing to both boundaries $x=\pm L$. Initial condition were $R_1(0)=1$, $R_2(0)=r/\nu-\epsilon$, $S_1(0)=0$, $S_2(0)=1$, where $\epsilon\sim 10^{-7}$. $\nu$ was determined by the shooting method to have $S_1(L)-S_1(-L)=6$, i.e., the third eigenfunction. The Dirac eigenfunctions can be computed analytically, however, we used the shooting method  with $S_2(0)=\pi$ and $S_1(L)-S_1(-L)=3\pi$.

Fig. \ref{fig:bound100} shows the 3-rd eigenfunctions of both equations for a box with $L=100$, which corresponds to the non-relativistic regime, meaning $\nu-r \ll r$. For the Dirac equation the energy eigenvalue equals $\nu=\sqrt{r^2+c^2k^2}\approx 1.00110971$, while for Eqs. \ref{fin:all} $\nu \approx 1.00104824$. Fig. \ref{fig:bound100}a shows that the amplitudes for both eigenfunctions are very similar and close to the non-relativistic solution of $\cos(kx)$, where $k=3\pi/2L$. Fig. \ref{fig:bound100}b shows that $S_1$ and $S_2$ of Eqs. \ref{fin:all} are very similar to $S_1/(\pi/2)$ and $S_2/(\pi/2)-1$ of Eqs. \ref{dirac2:all}. Thus, one can use analytical solutions of the Dirac equation to analyze the behavior of additive eigenvector variables. In particular, $\Delta S$ changes quickly between $\pm 1$ around the minima, and stays close to $0$ on the rest of the coordinate. Note, that while in Eq. \ref{fin:c} $\Delta S/\sqrt{1-\Delta S^2}$ turns to infinity as $\Delta S$ approaches $\pm 1$, its integral stays finite. This irregularity leads to non-smooth, cusp-like shapes for $R_i$ at the points. 

\begin{figure}[h]
\centering
\includegraphics[width=0.7\linewidth]{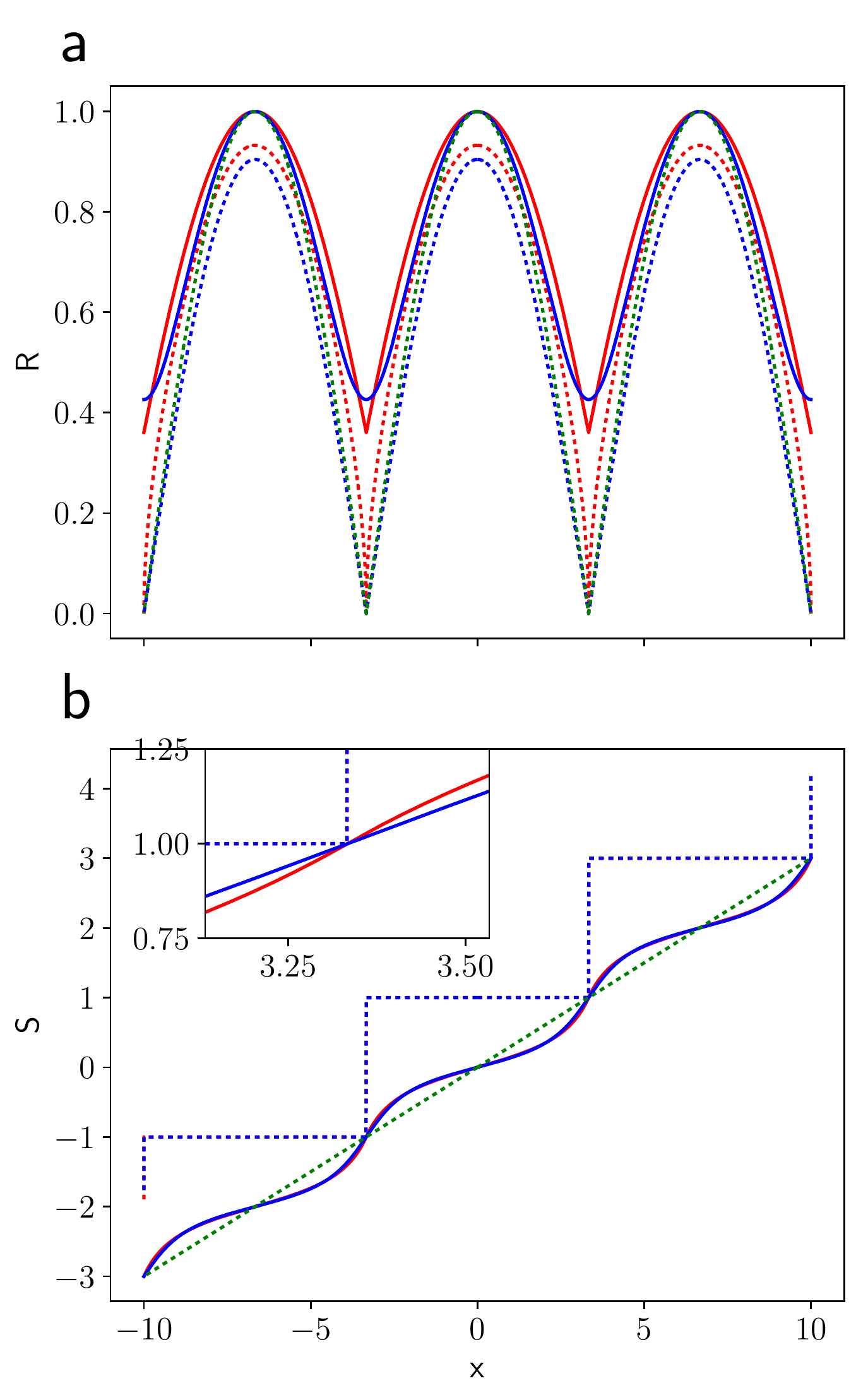}
\caption{3rd eigenfunctions of Eqs. \ref{fin:all} and \ref{dirac2:all} for $L=10$, where the relativistic effects become notable. Notations as in Fig. \ref{fig:bound100}}
\label{fig:bound10}
\end{figure}

Fig. \ref{fig:bound10} shows the 3-rd eigenfunctions of both equations for $L=10$ where relativistic effects become notable. The energy eigenvalue for the Dirac equation is
$\nu =\sqrt{r^2+c^2k^2}\approx 1.10547099$ while that for Eqs. \ref{fin:all} is $\nu \approx 1.07212227$. 

\begin{figure}[h]
\centering
\includegraphics[width=0.7\linewidth]{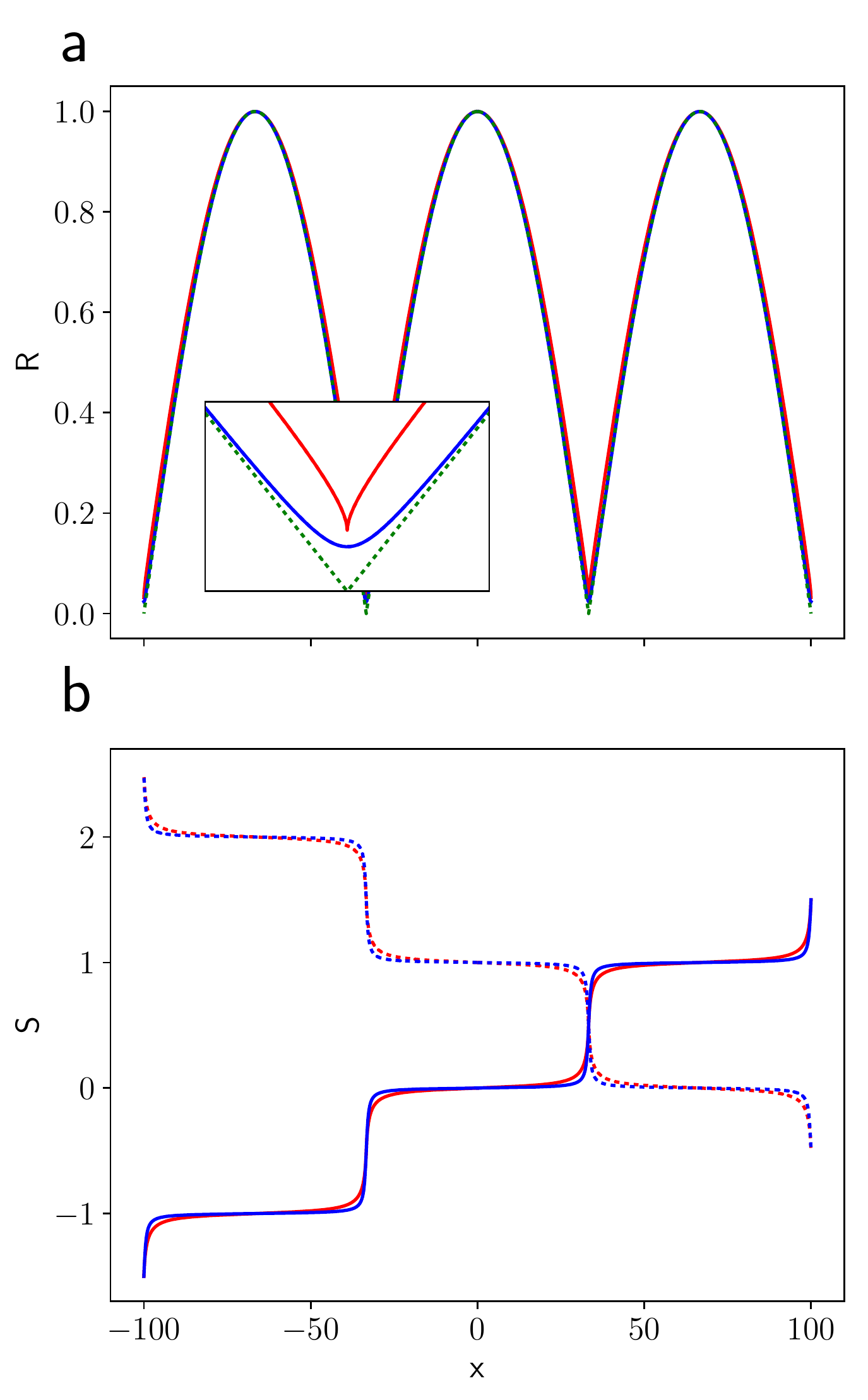}
\caption{3rd eigenfunctions of Eqs. \ref{fin:all} with $|S_2-S_1-1|>1$ and Eqs. \ref{dirac2:all} with $|S_2-S_1-\pi|>\pi$. Panel a) shows $R_1$ for Eqs. \ref{fin:all} and Eqs. \ref{dirac2:all} by red solid and blue solid lines, respectively. Non relativistic  eigenfunction $\cos(kx)$ is shown by green dotted line. Panel b) shows $S_1$ and $S_2$ of Eqs. \ref{fin:all} by solid and dotted red lines, respectively and $S_1/\pi$ and $S_2/\pi$ of Eqs. \ref{dirac2:all} by solid and dotted blue, respectively. The inset shows the magnified view of the region around the third local minimum.}
\label{fig:bound100_2}
\end{figure}

The Dirac equation also has solutions with $|S_2-S_1-\pi|>\pi$. For Eqs. \ref{fin:all}, one can obtain solutions with $|S_2-S_1-1|>1$, if one formally defines $\Delta S=(S_2-S_1)\bmod 2-1$, where$\mod 2$ defines the least positive reminder when divided by 2. Fig. \ref{fig:bound100_2} shows 3rd eigenfunctions for the box with $L=100$. We assumed that $R_1(x)=R_2(x)=R(x)$. 

\begin{figure}[htbp]
\centering
\includegraphics[width=0.9\linewidth]{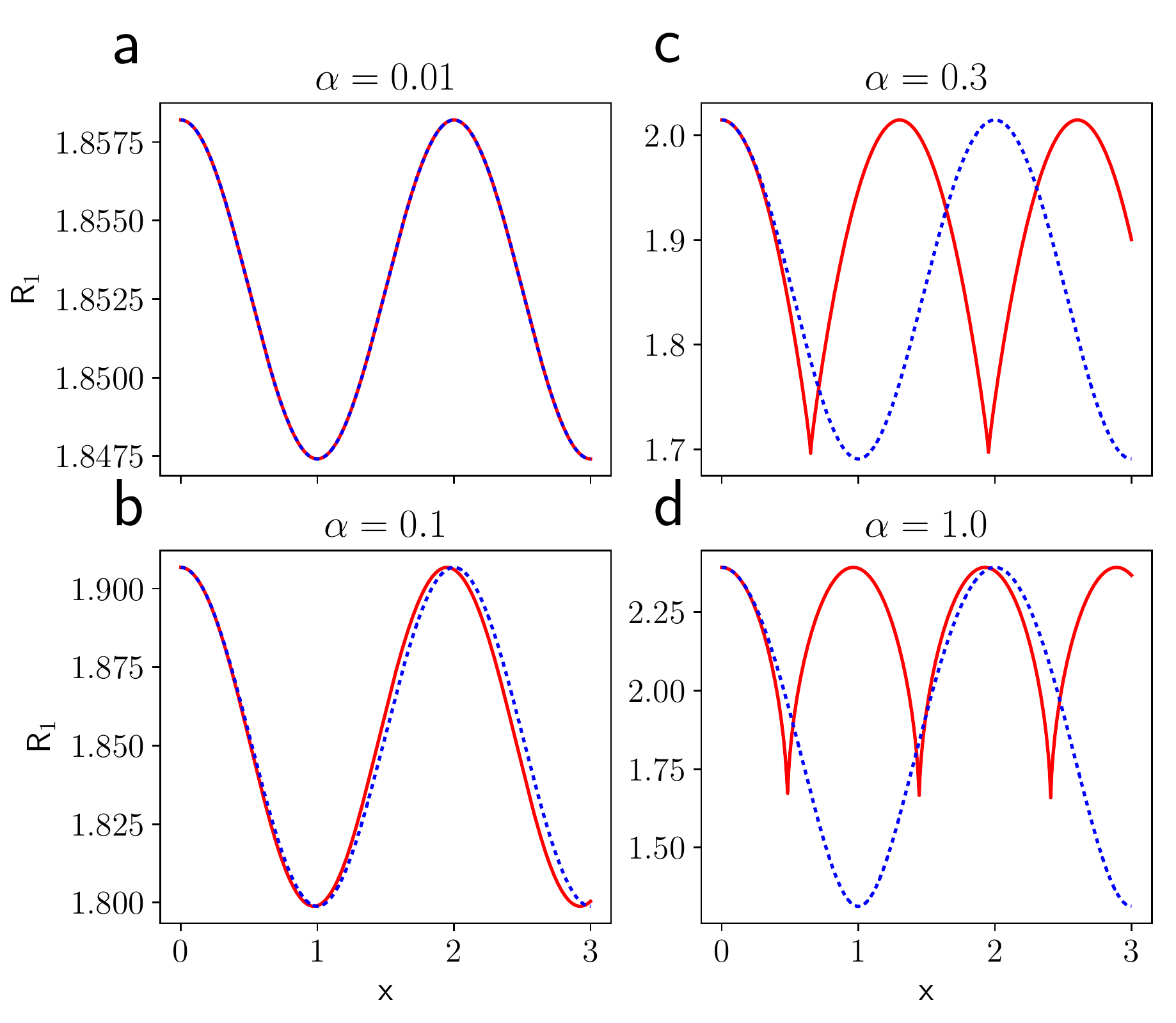}
\caption{Standing wave interference patterns obtained as a superposition of two eigenfunctions with opposite momenta, e.g., $\psi(k)+\alpha \psi(-k)$, for different values of $\alpha$. Solid red line shows $R_1$ for Eqs. \ref{fin:all} and dotted blue that for Eqs. \ref{dirac2:all}.}
\label{fig:free}
\end{figure}
Finally, we demonstrate an interesting property of Eqs. \ref{fin:all}, which is a consequence of its non-linearity. A superposition of two eigenfunctions of 
the Dirac equation with opposite momenta shows a standing wave pattern due to interference. By changing coefficients in the linear sum, one can change the shape of the wave, but not its wavelength - because of the linearity of the equations. In Eqs. \ref{fin:all} a sum of two solutions also shows a periodic standing wave pattern, 
but the wavelength now depends on the coefficients of the linear combination. 

To define a sum of two solutions for Eqs. \ref{fin:all} we proceed as follows. Starting with initial values $R_1(0)=\sqrt{\nu+ck}$, $R_2(0)=\sqrt{\nu-ck}$, $S_1(0)=0$ and $S_2(0)=1$ one can uniquely determine solution describing a wave moving to the right $R_1(x)$, $R_2(x)$, $S_1(x)$, $S_2(x)$. Starting with initial values $R_1(0)=\sqrt{\nu-ck}$, $R_2(0)=\sqrt{\nu+ck}$, $S_1(0)=0$ and $S_2(0)=1$ one can uniquely determine the solution describing a wave moving to the left. Starting with the linear combination of the initial values one can determine the solution corresponding to the combination of the solutions. Fig. \ref{fig:free} shows solutions obtained for different combinations of initial values $R_1(0)=\sqrt{\nu+ck}+\alpha \sqrt{\nu-ck}$, $R_2(0)=\sqrt{\nu-ck}+\alpha \sqrt{\nu+ck}$, $S_1(0)=0$, $S_2(0)=1$ in the relativistic regime ($\nu=\sqrt{r^2+c^2k^2}\approx 1.86$). For small $\alpha$ the difference between solutions of Eqs. \ref{fin:all} and the Dirac equations is small. As $\alpha$ increases the wavelength of the eigenfunctions of Eqs. \ref{fin:all} becomes shorter, while that of the Dirac equation stays the same. The shape of the eigenfunctions of Eqs. \ref{fin:all} changes as well by developing more pronounced cusps at the local minima. Solutions of Eqs. \ref{fin:all} shown on Fig. \ref{fig:free}.d have $|S_2-S_1-1|>1$ and we used $\Delta S=(S_2-S_1)\bmod 2-1$. In the non-relativistic regime, the non-linear effects (the wavelength shortening) are much smaller, as can be judged from the similarity of the eigenfunctions and eigenvalues.

\subsection{Comparison of solutions of Eqs. \ref{fin:all} with numerical simulations of conditioned Markov processes.}
Here we show that the derived equations (Eqs. \ref{fin:all} and \ref{tlgladdcond:all}) describe sub-ensembles of stationary trajectories, that can be obtained in a simulation. Note that the generation of such sub-ensembles is not a trivial task as one needs to simulate conditioned Markov processes. 

We first consider the simple case of position independent solutions which is relatively straightforward to simulate. The solutions can be interpreted as solutions with constant momentum analogous to the corresponding solutions of the Dirac equation. In this case stochastic trajectories are conditioned on the average speed $v/c=(P_1-P_2)/(P_1+P_2)=ck/\nu$. Namely, starting from a position $x$ and direction $d$ one simulates the telegraph process for time interval $T$. A trajectory is accepted if the total displacement $x(T)-x(0)$ equals $vT$, or more specifically if $|x(T)-x(0)-vT|<\epsilon$. If the trajectory is accepted, then simulation continues from a new position $x$, selected randomly in segment $[a,b]$ with uniform probability and with direction $d$ of the last point of the trajectory. If the trajectory is rejected, the simulation is continued from the last point and with the last direction of the last accepted trajectory.

\begin{figure}[htbp]
\centering
\includegraphics[width=0.7\linewidth]{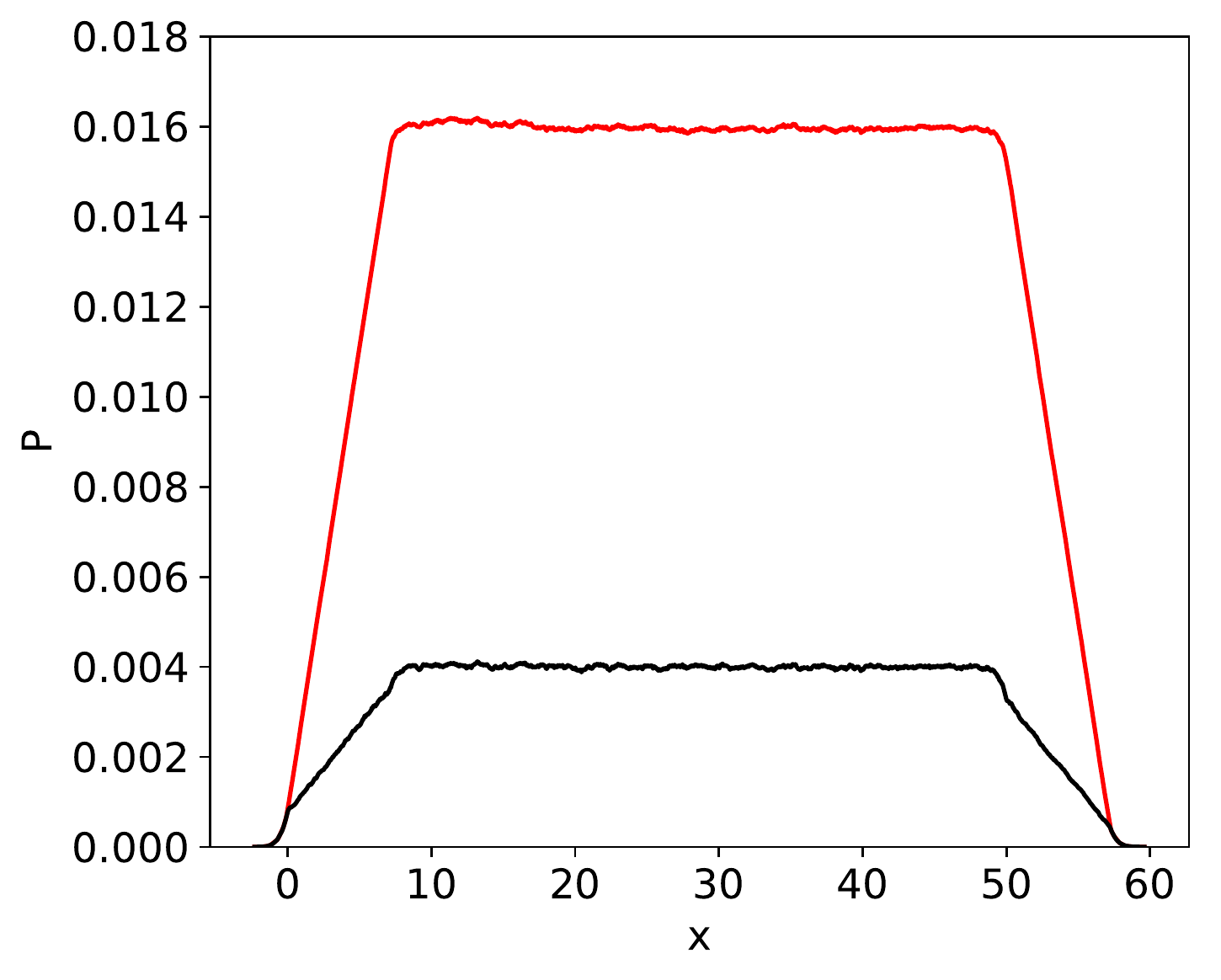}
\caption{Stationary trajectories conditioned on average speed. The red and black lines show $P_1(x)$ and $P_2(x)$, respectively. }
\label{fig:cond_v}
\end{figure}

Note that the numerical efficiency of the algorithm depends crucially on the parameters. In particular, it decreases exponentially with time interval $T$.
Selecting small $\epsilon$ increases the accuracy, however decreases the efficiency of the algorithm, as many trajectories with small $|x(T)-x(0)-vT|$ are rejected. The following parameters were used: $r=1$, $c=1$, $T=12$, $v/c=0.6$, $a=0$, $b=50$, $\epsilon=0.01$. Fig. \ref{fig:cond_v} shows $P_1(x)$ and $P_2(x)$. The probabilities are constant (apart from the boundary effects) and are in agreement with the equations; namely $P_1/P_2=(1+v/c)/(1-v/c)=4$. The transition probabilities of the conditioned Markov process, estimated from the trajectories, are also in agreement with the equations (Eq. \ref{tlgladdcond:all}b). Namely, for rate $rq_2/q_1=2$ direct estimate from the trajectory gives $2.1$.

Simulation of stationary solutions that mimic a free particle in a box (Fig. \ref{fig:bound100}) is more complicated, as we have not yet developed a direct simulation algorithm for additive eigenvectors. One can use the QSD formalism and simulate the conditioned Markov process by using the standard Markov process with modified rate matrix (Eqs. \ref{kij}). However, it means that one first has to solve the equations to find $q(i)$. We will use an indirect algorithm. The probability distribution on Fig \ref{fig:bound100}a suggests that a particle can be roughly described as to be confined to a segment. Which suggests the idea to model this stationary solution by an ensemble of stationary trajectories that are conditioned to stay inside a segment. We first consider the non-relativistic case, where one may assume with good accuracy that
probability for trajectories to be found at the segment boundaries is zero. 

The specific algorithm is as follows.  Starting from a point on a segment $[a,b]$, one simulates the telegraph process for sufficiently long time $T$. If the trajectory during this simulation has stayed in the segment, it is accepted, and the process is continued from the last point of the trajectory. If the trajectory has left the segment, it is rejected, and the process is repeated from the last point of the last accepted trajectory. In such a way we keep only trajectories which stay inside the segment, i.e., stochastic periodic trajectories in some sense. The longer is the time interval, the more "revolutions" inside the segment the trajectory will make. The method is highly inefficient, because the probability to stay in the segment decreases exponentially with time $T$, i.e., only exponentially small number of trajectories are accepted. 

\begin{figure}[htbp]
\centering
\includegraphics[width=0.7\linewidth]{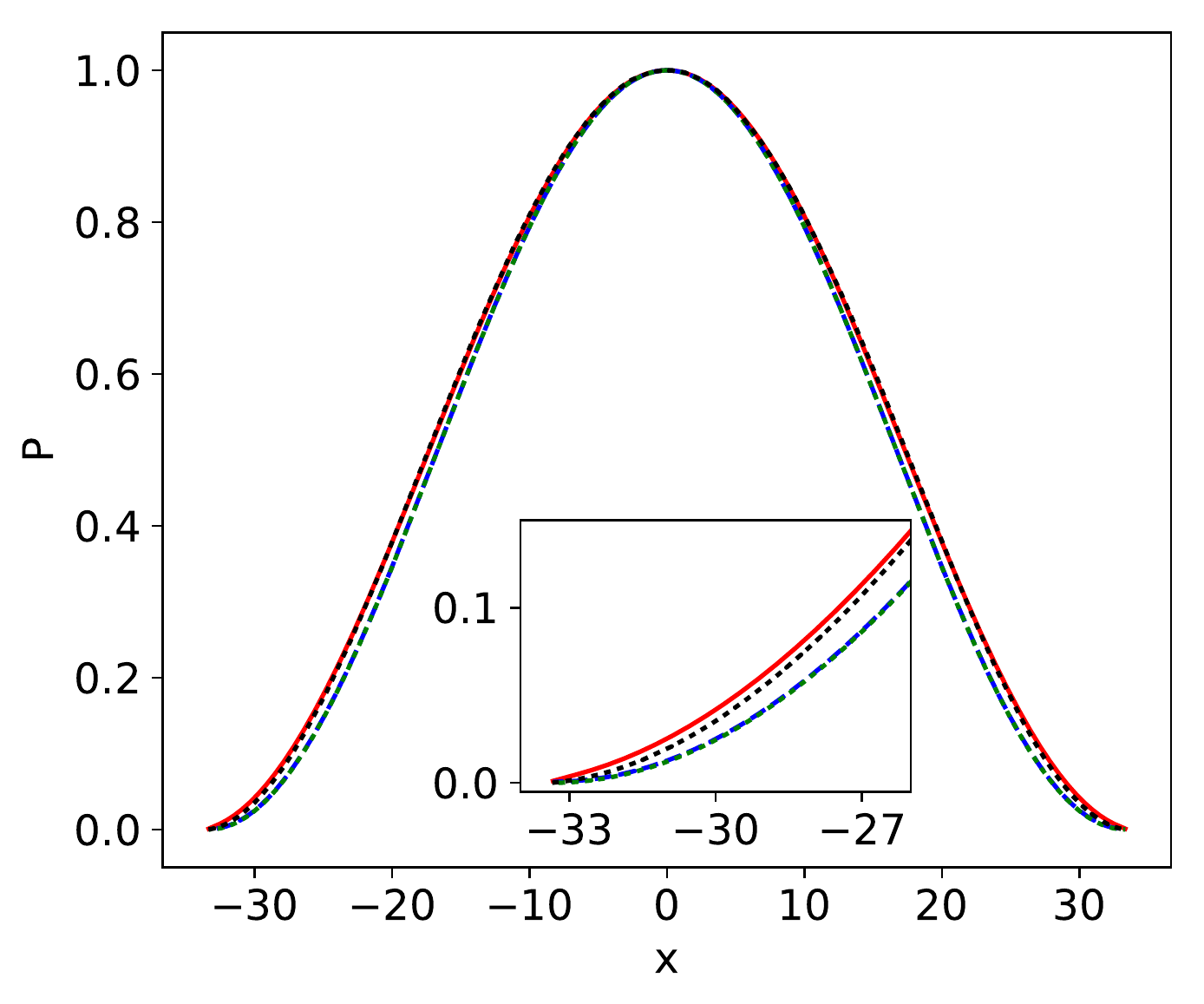}
\caption{Stationary trajectories conditioned on staying inside a segment - comparison of probabilities. The black dotted line shows the simulation results, the red solid line shows solution of Eqs. \ref{fin:all}, the blue dashed line shows solution of the Dirac equation, the green dotted line shows non-relativistic solution $P(x)=\cos^2(3\pi/200 x)$. The last two lines are practically indistinguishable. The inset shows the magnified view of the left corner.}
\label{fig:segm}
\end{figure}

In order to simulate the solution shown on Fig. \ref{fig:bound100} the following parameters were selected $r=1$, $c=1$, $a=-100/3$, $b=100/3$, $T=3000$. Fig. \ref{fig:segm} shows that the probabilities obtained in the simulation are in excellent agreement with those computed from the equations. The magnified view shows that the simulation results are closest to the solution of the derived equations (Eqs. \ref{fin:all}). The solutions were obtained assuming that $R_1(x)=R_2(x)$.

\begin{figure}[htbp]
\centering
\includegraphics[width=0.7\linewidth]{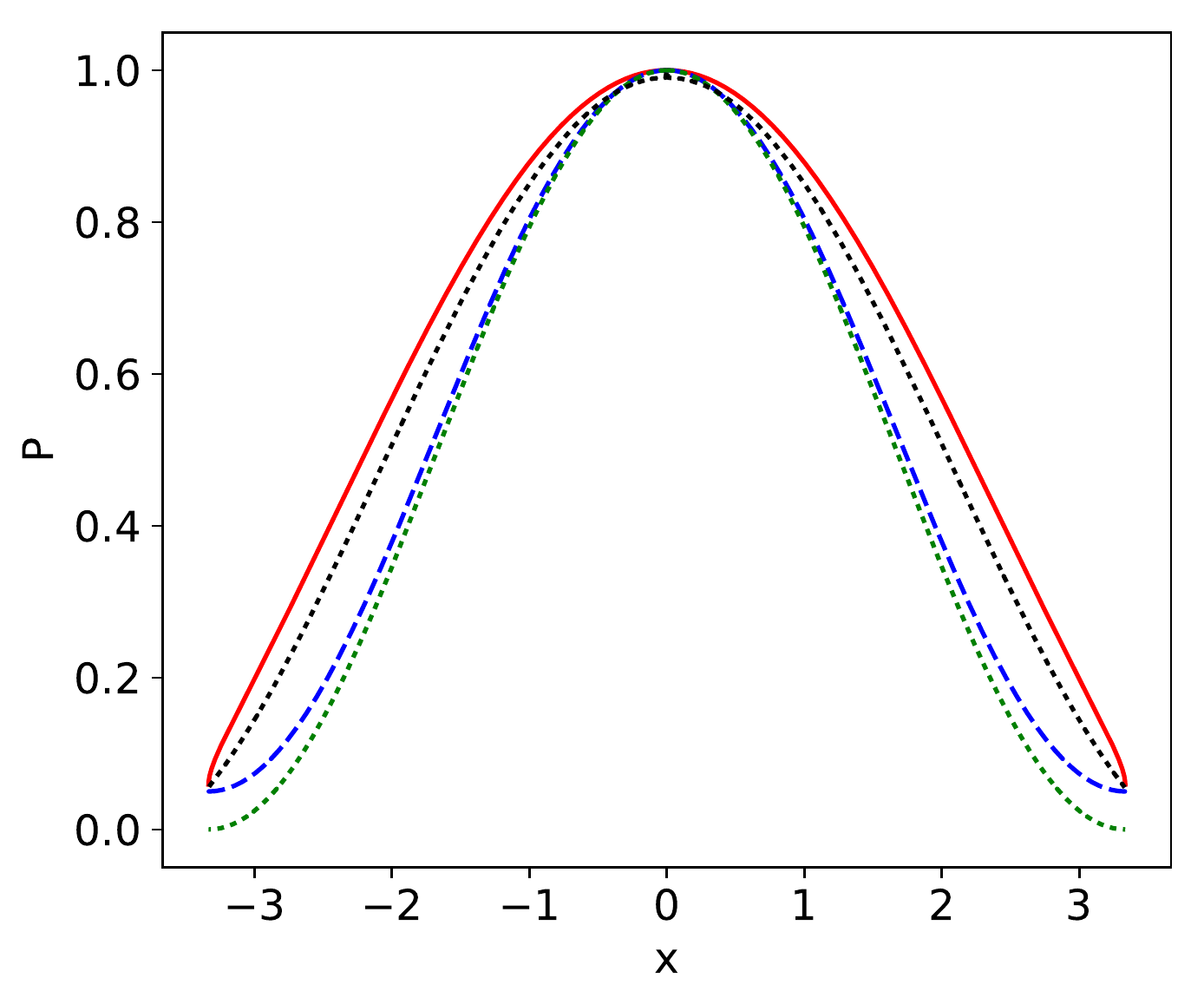}
\caption{Stationary trajectories conditioned on staying inside a segment in the relativistic regime - comparison of probabilities. Notations as in Fig. \ref{fig:segm}.}
\label{fig:segmr}
\end{figure}

To simulate trajectories corresponding to the solutions in the relativistic regime (Fig. \ref{fig:bound10}) the algorithm requires the following modification. The solutions of both the derived and the Dirac equations have non-zero probabilities on the boundaries of the segment. It means that trajectories are passing through the boundaries. In the non-relativistic regime these probabilities are very small and can be safely neglected. To mimic non-zero probabilities at the boundaries, we allowed the simulated trajectory to pass through the boundaries with some probability $p_c$. This probability $p_c$ was determined by fitting the boundary probabilities. In order to simulate the solution shown on Fig. \ref{fig:bound100} the following parameters were selected $r=1$, $c=1$, $a=-10/3$, $b=10/3$, $T=100$, $p_c=0.23$. Fig. \ref{fig:segmr} shows that the probabilities obtained in the simulation are in good agreement with those computed from the equations, considering the shortcomings of the algorithm. Again, the best agreement is with the solution of the derived equations. The solutions were obtained assuming that $R_1(x)=R_2(x)$.

\section*{Concluding Discussion}
We have suggested a theory of additive eigenvectors for the description of stochastic Markov processes. Such eigenvectors appear when one uses additive separation ansatz $S(i,t)=W(i)-\nu t$ instead of standard multiplicative ansatz $P(i,t)=u(i)e^{\mu t}$. Additive eigenvectors have very peculiar properties and provide description of stochastic dynamics very different from the conventional one. In particular, an additive eigenvector describes a conditioned Markov process, which consists of a sub-ensemble of stationary trajectories, conditioned to have the same additive eigenvectors for forward and time-reversed dynamics. This fact was used to derive equations for additive eigenvectors for a continuous-time Markov chain (Eqs. \ref{condaddev}-\ref{condaddevt}). In contrast to conventional eigenvectors, each additive eigenvector describes a stationary stochastic process with its own stationary probability distribution. This process is periodic in the sense that it is described 
by an additive eigenvector, which describes an advancement of the phase.

The formalism was illustrated by considering the stochastic telegraph process. Differential equations, describing additive eigenvectors, have been derived. The derived equations are structurally similar to the one-dimensional Dirac equations and for some values of parameters they have similar solutions, e.g., in the non-relativistic regime. This, in particular, confirms that an additive eigenvector describes a stochastic process periodic in the conventional sense. It also shows that these processes have wave properties and can exhibit interference effects. An important difference, however, is that the derived equations are non-linear. We emphasize that the derived equations describe a classical stochastic process. Numerical simulations of conditioned stationary trajectories of the telegraph process confirmed the validity of the formalism.

Let's discuss how the additive eigenvectors can be used to describe stochastic dynamics. Consider, again, a long systems trajectory describing diffusion on Fig. \ref{fig:twominima} and assume that the first additive eigenvector describes the TP ensemble, while the second additive eigenvector describes dynamics inside the basins. Project the trajectory on the first additive eigenvector, i.e., compute the time-series of the phase $W$. Since the phase is multivalued it is easier to just add up the increments, e.g.,  $W(i)-W(j)$ for a transition from $j$ to $i$. The phase will increase with time with average speed $\nu$ indicating that the system continues the periodic movement between the basins. The same trajectory, projected on the second additive eigenvector should analogously describe periodic movement inside the basins. And analogously for all other additive eigenvectors. In such a way one can decompose stochastic dynamics into a collection of independent stochastic periodic motions - stochastic eigenmodes. 

This suggests the additive eigenvectors as a new type of optimal reaction coordinates \cite{KrivovMethoddescribestochastic2013, BanushkinaOptimalreactioncoordinates2016}. The additive eigenvectors are free of shortcomings of both the standard eigenvectors and the committor functions, which are often used as such coordinates. A problem with the committor functions, is that they require specification of two boundary states. First, the specification of such states for complex systems is a difficult task. Second, it means that the committor accurately describes only the transition dynamics between the states. The dynamics inside the states is not described at all. The eigenvectors do not need boundary states, however, as was explained in the introduction, their contribution to the description of stationary dynamics decrease exponentially with time. Thus, the additive eigenvectors, which do not require boundary states and are stationary are good candidates for optimal reaction coordinates.


Given the decomposition of dynamics on additive eigenvectors, one should be able to reconstruct the stochastic dynamics back. For example, if one considers dynamics, described by a single additive eigenvector, one can use $q(i)$ to determine the rates of the Markov process that specifies the dynamics (Eq. \ref{kij}). For a general solution describing a superposition of additive eigenvectors, $q(i)$ and, correspondingly, the rates of the Markov process shall be time dependent. It is an  illustration of interference effects, i.e., the stationary probability for a superposition of additive eigenvector is not equal to the superposition of corresponding stationary probabilities. If one combines all the additive eigenvectors, then one should be able to reconstruct the original Markov dynamics, if the set of additive eigenvectors is complete, or only a part of it, if otherwise. In the later case, a Markov process shall have a generic decomposition on a process described by additive eigenvectors and an orthogonal process. 

Consider an ensemble of weakly interacting subsystems, which can be considered as independent. It is straightforward to show (see Appendix \ref{ensemble}) that an additive eigenvector for the ensemble  can be expressed from those for the subsystems as $W=\sum_{\alpha} W_{\alpha}$, $q=\prod_{\alpha} q_{\alpha}$, $v=\prod_{\alpha} v_{\alpha}$ and $\nu=\sum_{\alpha} \nu_{\alpha}$, where $\alpha$ denotes subsystems. The last equation, which is equivalent to the usual $E=\sum_\alpha E_\alpha$, corresponds to the microcanonical ensemble. Thus the formalism suggests that the additive eigenvectors in subsystems should be distributed according to the corresponding statistical distribution, e.g., Boltzmann or possibly Fermi-Dirac, taking into account the similarity between the equations.


A lot of research has been devoted to find a stochastic model/interpretation of quantum mechanical equations, e.g., \cite{Schrodinger1931, NelsonDerivationSchrodingerEquation1966, GaveauRelativisticExtensionAnalogy1984, NagasawaStochasticProcessesQuantum2000} to mentioned just a few works, most relevant to our work. The main difference of this work is the usage of additive eigenvectors and straightforward, self-contained stochastic interpretation - one can completely avoid any reference to quantum mechanics. Relativistic quantum mechanical equations appeared here because they have similar properties. Still, our results illustrate how one can derive a Dirac-like equation from the telegraph process without using an analytic continuation, e.g., $t\rightarrow it$ \cite{GaveauRelativisticExtensionAnalogy1984}. A disadvantage of the analytic continuation is that it is difficult to interpret the obtained equations as the time variable loses its conventional meaning. Our Eq. \ref{tlgladdcond:all} has a straightforward stochastic interpretation at the expense of being not exactly equal to the Dirac equation and being non-linear. The difference between the two equations is of the order $O(\Delta S^3)$ and thus is very small for small $\Delta S$.

We have illustrated the formalism of additive eigenvectors  by considering the telegraph process. An important task is to extend the formalism to the description of diffusive dynamics. In particular, one-dimensional diffusion, where the transition probabilities are distributed as $P(x,y,\Delta t)=e^{-\frac{(x-y)^2}{2D \Delta t}}$ and $D$ is the diffusion coefficient. Note, that Eq. \ref{fin:all}, in the non-relativistic regime of very large $r$, can be used for an approximate description of one-dimensional diffusion. To extend our formalism to the diffusion one has to do the following. First, one needs to specify the multivalued character of the additive eigenvectors. Since the diffusion satisfies the detailed balance, additive eigenvectors are necessarily multivalued. An analogous approach assuming that displacements to the right ($x-y>0$) and to the left ($x-y<0$) belong to different branches leads to the following problem: infinitesimally small displacements around $0$, which are infinitesimally close to each other, belong to different branches. Second, one needs to derive equations analogous to Eqs. \ref{condaddev} for the diffusion process. One way to solve both these problems, is to consider the diffusion as a large scale description of the random walk. Consider the telegraph process as a model for the random walk. Large scale description means that large scale change $\Delta S$ is composed of many small changes $\Delta S_i$ during many steps $1 \rightarrow 2 \rightarrow 1 ... \rightarrow 2$, where $1$ and $2$ denote movement to the right and left, respectively. If all $\Delta S_i\ll 1$ then equations for additive eigenvectors for the telegraph process become equal to the Dirac equation. Thus it is straightforward to conjecture that additive eigenvectors for the diffusion on the line are described by the one-dimensional Dirac equation. Equations for additive eigenvectors for the three dimensional diffusion should be linear and of first order and should  respect rotational invariance as well as the Lorentz invariance. Thus we conjecture that (some) additive eigenvectors for the diffusion in three dimensions are described by the three dimensional Dirac equation. Obviously, the conjecture also suggests that the Dirac equation is not an exact fundamental equation but rather an approximate large scale description similar to the diffusion equation.


\section{Appendix}

\subsection{Properties of additive eigenvectors}
\label{propev}
Here we show that the fact that additive eigenvectors are multivalued is quite generic. Consider Eqs. \ref{addevgen}. It is easy to show that if both equations are valid for $\Delta t=\Delta t_0$, then they are valid for arbitrary large time intervals $\Delta t=k \Delta t_0$, as it should be for an eigenvector. By pulling $W(i)$ out of summation sign one sees that 
\begin{align}
\nu \Delta tP^{st}(i)=\sum_j P_{\Delta t}(i|j)P^{st}(j)(W(i)-W(j))
\label{addevgen2}
\end{align}
equals Eq. \ref{addevgen}b.
For a system with bounded configuration space, i.e., limited number of states, the right hand side is bounded by $\max |W(i)-W(j)|$ if $W(i)$ is singlevalued. By considering the limit of $\Delta t \rightarrow \infty $ it follows that $\nu =0$. Thus, to have $\nu\ne 0$, $W(j)$ should be multivalued.

While the fact that additive eigenvectors are multivalued may not seem very surprising, in some limiting cases it may lead to extremely counter-intuitive properties. Consider, how the equations on additive eigenvector 
for the model system, $W(j)-W(j-1)=\nu/r$, look in the limiting case of just two states $N=2$:
\begin{subequations}
\begin{align}
&W(1)-W(0)=\nu/r\\
&W(0)-W(1)=\nu/r,
\end{align}
\label{mv0}
\end{subequations}
i.e., 
\begin{align}
W(1)-W(0)+W(0)-W(1)=2\nu/r\ne 0.
\label{mv1}
\end{align}

By summing Eq. \ref{addevgen2} over $i$ one obtains 
\begin{align}
\nu \Delta t=\sum_{ij} P_{\Delta t}(i|j)P^{st}(j)(W(i)-W(j)).
\label{addevgen3}
\end{align}
For a system with the detailed balance $P_{\Delta t}(i|j)P^{st}(j)=P_{\Delta t}(j|i)P^{st}(i)$ the right side is exactly zero as $W(i)-W(j)$ is antisymmetric, while $P_{\Delta t}(i|j)P^{st}(j)$ is symmetric. Again, solutions with $\nu\ne 0$ are possible only if for some $i$ and $j$ 
$W(i)-W(j)+W(j)-W(i)\ne 0$, which is possible if $W$ is multivalued (see Eqs. \ref{mv0} and \ref{mv1}). Thus, in order to obtain non-trivial solutions with $\nu\ne 0$ in practically important cases of finite configuration space or dynamics with the detailed balance, the additive eigenvectors should be multivalued.

Let us obtain an expression for $\nu$. In the limit of large $\Delta t$ the right hand side of Eq. \ref{addevgen3} can be estimated as $\Delta W N_c$, if one considers a single long (ergodic) trajectory and assumes that $W$ is multivalued; here $\Delta W$ is the increment of the additive eigenvector for one cycle and $N_c$ is the number of cycles. For normalization $\Delta W=2\pi$ one obtains $\nu= 2\pi/\tau$, where $\tau=\Delta t/N_c$ is the mean time to complete the cycle.

\subsection{Derivation of Eqs. \ref{tlgladdcond:all}.}
\label{dertlg}
Here we derive differential equations for additive eigenvectors for the telegraph process. We use the following Markov chain that approximates the telegraph process.
System at position $i \Delta x$ in internal state $1$ (which describes movement to the right) can move to positions $(i+1)\Delta x$ with rate $a=c/\Delta x$ and can move to the left, to the position $(i-1)\Delta x$ and internal state $2$, with rate r. And correspondingly, system at position $i \Delta x$ in internal state $2$ (which describes movement to the left) can move to positions $(i-1)\Delta x$ with rate $a$ and can move to the right, to the position $(i+1)\Delta x$ and internal state $1$, with rate r. Thus the rate matrix reads $K^0(1,i+1|1,i)=a$, $K^0(2,i-1|1,i)=r$, $K^0(2,i-1|2,i)=a$, $K^0(1,i+1|2,i)=r$.
Eq. \ref{condaddevt}d for $P(1,i)$ reads 
\begin{align*}
\partial P(1,i)&/\partial t=K^0(1,i|1,i-1)q(1,i)/q(1,i-1)P(1,i-1) \nonumber\\
&+K^0(1,i|2,i-1)q(1,i)/q(2,i-1)P(2,i-1)\nonumber\\
&-K^0(1,i+1|1,i)q(1,i+1)/q(1,i)P(1,i)\nonumber\\
&-K^0(2,i-1|1,i)q(2,i-1)/q(1,i)P(1,i)
\end{align*}
or
\begin{align*}
\partial P(1,i)/\partial t+c/\Delta x[&q(1,i)/q(1,i-1)P(1,i-1) \nonumber\\
-&q(1,i+1)/q(1,i)P(1,i)]=\nonumber\\
rq(1,i)/q(2,i-1)&P(2,i-1)\nonumber\\
-r&q(2,i-1)/q(1,i)P(1,i)
\end{align*}
which, in the limit of $\Delta x \rightarrow 0$ becomes
\begin{align*}
\partial P_1/\partial t_+=r q_1/q_2P_2-rq_2/q_1P_1
\end{align*}
where $\partial /\partial t_+=\partial /\partial t+c\partial /\partial x$ and we use the shortened notation, e.g., $P_1(x,t)=P(1,x,t)$ and $P_2(x,t)=P(2,x,t)$, and similar for the other quantities.

Eq. \ref{condaddevt}c for $S(1,i)$ reads 
\begin{align*}
&c/\Delta xq(1,i+1)/q(1,i)[S(1,i+1)-S(1,i)]\\
&+rq(2,i-1)/q(1,i)[S(2,i-1)-S(1,i)]=-\partial S(1,i)/\partial t
\end{align*}
which, in the limit of $\Delta x \rightarrow 0$ becomes
\begin{align*}
\partial S_1/\partial t_+ +rq_2/q_1(S_2-S_1)=0.
\end{align*}
Other equations are derived analogously.

\subsection{Derivation of Eqs. \ref{fin:all} using Lorentz transformations.}
\label{derlor}
Here we show how Eqs. \ref{fin:all} can be derived by considering Lorentz transformations instead of conditioned Markov processes. In this section we consider only conventional (conditioning-free) Markov processes, which have a unique stationary distribution, and consequently, a single stationary additive eigenvector, which we call the rest frame additive eigenvector. Other stationary additive eigenvectors with different stationary distributions are interpreted as the rest frame eigenvector observed in different frames of reference.

For reader convenience we copy here Eqs. \ref{tlg:all} describing telegraph process
\begin{subequations} \label{tlg2:all}
\begin{align}
\partial P_1/ \partial t_+&= -r_{21} P_1 +r_{12} P_2\\
\partial P_2/ \partial t_-&= -r_{12} P_2 + r_{21} P_1,
\end{align}
\end{subequations}
Equations for additive eigenvectors (Eqs. \ref{addevgenall}) for forward and time reversed stationary dynamics $P^{st}_1=P^{st}_2=1$ of the telegraph process with $r_{12}=r_{21}=r$ are
\begin{subequations}
\begin{align}
&\partial S_1/\partial t_+ +r[S_1-S_2]=0\\
&\partial S_2/\partial t_- +r[S_2-S_1]=0\\
&\partial S_1/\partial t_+ +r[S_2-S_1]=0\\
&\partial S_2/\partial t_- +r[S_1-S_2]=0
\end{align}
\end{subequations}
Introducing $S_2-S_1=1+\Delta S$ and $S_1-S_2=1-\Delta S$, one finds single trivial solution $S_1=-\nu t$, $\Delta S=0$, which is the solution in the rest frame.

\subsubsection{Global Lorentz transformations.}
\label{sect:gl}
Let us mark the variables describing the above solution in the rest frame with primes (e.g., $t',x', t'_+, r'_{12}$ ). Consider Lorentz transformation to the moving frame ($t, x$): $dt=\frac{dt'-vdx'/c^2}{\sqrt{1-v^2/c^2}}$ and $dx=\frac{dx'-vdt'}{\sqrt{1-v^2/c^2}}$, written in light cone coordinates $d t'_+=dt_+a $ and $d t'_-=dt_-/a $, where $a=\frac{\sqrt{1+v/c}}{\sqrt{1-v/c}}$. Here we assume that $v$ and $a$ are constant, i.e., the Lorentz transformation is global. Then $r_{21}=r'_{21}a=ra$, $r_{12}=r'_{12}/a=r/a$.  The equations for probability (Eqs. \ref{tlg2:all}) become
\begin{subequations} \label{tlga:all}
\begin{align}
\partial P_1/ \partial t_+&= -ra P_1 +r/a P_2\\
\partial P_2/ \partial t_-&= -r/a P_2 + ra P_1,
\label{st}
\end{align}
\end{subequations}
with (non-normalized) stationary probabilities $P^{st}_1=1/a$ and $P^{st}_2=a$; $a=\sqrt{P_2^{st}/P_1^{st}}=R_2/R_1$.
Equations for additive eigenvectors for forward and time reversed dynamics are:
\begin{subequations}\label{addevtlg0}
\begin{align}
\partial S_1/\partial t_+ + [S_1-S_2]ra&=0\\
\partial S_2/\partial t_- + [S_2-S_1]r/a&=0\\
\partial S_1/\partial t_+ + [S_2-S_1]ra&=0\\
\partial S_2/\partial t_- + [S_1-S_2]r/a&=0.
\end{align}
\end{subequations}
Assuming $S_1=kx-\nu t$, $S_2-S_1=1+\Delta S$, $S_1-S_2=1-\Delta S$, one finds $\Delta S=0$, $\nu^2= r^2+c^2k^2$, and $a=R_2/R_1=\sqrt{\frac{\nu-ck}{\nu+ck}}$.

\subsubsection{Local Lorentz transformations.}
Global Lorentz transformation, considered in the previous section, allowed us to obtain position independent solutions. Can one obtain solutions with non-constant $R_i$? Generalizing the above one may expect that position dependent $P_i^{st}$ can be obtained with local (position dependent) Lorentz transformations $dt'_+=dt_+a(x)$ and $dt'_-=dt_-/a(x)$. Note that straightforward $a(x)=R_2(x)/R_1(x)$ leads to $\partial P_1^{st}/\partial t_+=\partial P_2^{st}/\partial t_-=0$, which for the stationary case means $P_i^{st}=const$. One needs to solve the inverse problem of finding such position dependent transformations i.e., $a(x)$, which reproduce the desired $P_i^{st}(x,t)$. To this end one solves Eqs. \ref{tlga:all} and finds
\begin{subequations} \label{aa2:all}
\begin{align}
a&=R_2/R_1(\sqrt{1+\Delta^2} -\Delta) \\
r_{21}&=rR_2/R_1(\sqrt{1+\Delta^2} -\Delta) \\
r_{12}&=rR_1/R_2(\sqrt{1+\Delta^2} +\Delta)\\
\Delta&=\frac{1}{rR_2}\frac{\partial R_1}{\partial t_+}=-\frac{1}{rR_1}\frac{\partial R_2}{\partial t_-}
\end{align}
\end{subequations}
Stochastic dynamics, described by Eq. \ref{tlg2:all} with such rates, can have any specified stationary probability distributions $P_1^{st}$ and $P_2^{st}$ (which, obviously, should satisfy $\partial P_1^{st}/ \partial t_+ + \partial P_2^{st}/\partial t_-=0$). Another interpretation is that stochastic dynamics in the rest frame $(t',x')$, described by Eq. \ref{tlg2:all} with constant rates $r'_{12}=r'_{21}=r$, where  $P_1^{st'}$ and $P_2^{st'}$ are constant and equal, shall have desired $P_1^{st}$ and $P_2^{st}$ in $(t,x)$ frame,
which is obtained by local Lorentz transformation with corresponding $a(x)$, i.e., $dt'_+=dt_+a(x)$ and $dt'_-=dt_-/a(x)$. Intuitively, a varying probability distribution results from the fact that the system spends less time moving in one direction than the other, thus it makes smaller steps in one direction than the other, which leads to a stationary distribution that is position dependent.

Comparing Eqs. \ref{tlga:all} with Eqs. \ref{tlgladdcond:all}b,c one sees that rescaling of rates due to additive eigenvector conditioning is equivalent to that due to Lorentz transformations ($a=q_2/q_1$).

To find which transformations ($a(x)$) represent an additive eigenvector, one applies Eqs. \ref{addevgenall} to Eqs. \ref{tlg2:all} with the corresponding $r_{21}$ and $r_{12}$ and obtains
\begin{subequations} \label{prefin}
\begin{align}
\partial S_1/\partial t_+ +[S_1-S_2]rR_2/R_1(\sqrt{1+\Delta^2} +\Delta)&=0\\
\partial S_2/\partial t_- +[S_2-S_1]rR_1/R_2(\sqrt{1+\Delta^2} -\Delta)&=0\\
\partial S_1/\partial t_+ +[S_2-S_1]rR_2/R_1(\sqrt{1+\Delta^2} -\Delta)&=0\\
\partial S_2/\partial t_- +[S_1-S_2]rR_1/R_2(\sqrt{1+\Delta^2} +\Delta)&=0
\end{align}
\end{subequations}
Introducing $S_2-S_1=1+\Delta S$ and $S_1-S_2=1-\Delta S$, one obtains that $\Delta S=\Delta /\sqrt{1+\Delta ^2}$, $\Delta =\Delta S/\sqrt{1-\Delta S^2}$ and $a(x)=R_2/R_1\sqrt{\frac{1-\Delta S}{1+\Delta S}}$. The equations are simplified to Eqs. \ref{fin:all}.

\subsection{Additive eigenvector for an ensemble for independent subsystems}
\label{ensemble}
Here we show that additive eigenvector for an ensemble of independent subsystem can expressed from those for the subsystems as
\begin{subequations} \label{ens1}
\begin{align}
W(i)&=\sum_\alpha W_\alpha(i_\alpha)\\
\nu&=\sum_\alpha \nu_\alpha\\
q(i)&=\prod_\alpha q_\alpha(i_\alpha)\\
v(i)&=\prod_\alpha v_\alpha(i_\alpha),
\end{align}
\end{subequations}
here $\alpha$ denotes subsystem, $i_\alpha$ denote states in subsystem $\alpha$ and $i=(i_1,...,i_\alpha,...)$ denotes states in the ensemble, which is a Cartesian product of states of subsystems. 

We consider just an ensemble of two subsystems, for brevity; generalization to larger number is straightforward. Independence of subsystems means that transition probability for an ensemble equals product of transition probabilities for subsystems: $P^0(i_1,i_2|j_1,j_2)=P^0_1(i_1|j_1)P^0_2(i_2|j_2)$. Using approximation $P^0(i_1,i_2|j_1,j_2)=e^{\Delta t K^0(i_1,i_2|j_1,j_2)}\approx \delta_{i_1 j_1}\delta_{i_2 j_2}+\Delta t K^0(i_1,i_2|j_1,j_2)$, and analogous for the subsystems, and by taking limit $\Delta t\rightarrow 0$ one arrives at the expression for the reaction rate matrix for an ensemble of two independent subsystems.
\begin{align} \label{ens2}
K^0(i_1,i_2|j_1,j_2)=\delta_{i_1 j_1}K^0_2(i_2|j_2)+ \delta_{i_2 j_2}K^0_1(i_1|j_1)
\end{align}
Entering these quantities into Eq. \ref{condaddev} one finds that they satisfy these equations. In particular for Eq. \ref{condaddev}c one has  
\begin{align}
&\sum_{i_1,i_2}K^0(i_1,i_2|j_1,j_2)\frac{q(i_1,i_2)}{q(j_1,j_2)}[W(i_1,i_2)-W(j_1,j_2)]=\nonumber\\
&\sum_{i_1}K^0_1(i_1|j_1)\frac{q_1(i_1)}{q_1(j_1)}[W_1(i_1)-W_1(j_1)]+\\
&\sum_{i_2}K^0_2(i_2|j_2)\frac{q_2(i_2)}{q_2(j_2)}[W_2(i_2)-W_2(j_2)]=\nonumber \nu_1+\nu_2=\nu \nonumber
\end{align}


\end{document}